\documentclass[AMA,STIX1COL,table,hyperfootnotes=false]{WileyNJD-v2}
\usepackage{float}
\floatstyle{plaintop}
\restylefloat{table}
\usepackage{moreverb}
\usepackage{url}
\usepackage{amsmath}
\usepackage{listings}
\usepackage{endnotes}
\usepackage{subcaption}
\usepackage{xcolor}
\usepackage{graphicx}
\usepackage{hyperref}
\usepackage[binary-units]{siunitx}
\usepackage{tikz}

\newcommand\BibTeX{{\rmfamily B\kern-.05em \textsc{i\kern-.025em b}\kern-.08em
T\kern-.1667em\lower.7ex\hbox{E}\kern-.125emX}}
\articletype{RESEARCH ARTICLE}
\received{<day> <Month>, <year>}
\revised{<day> <Month>, <year>}
\accepted{<day> <Month>, <year>}

\definecolor{codegreen}{rgb}{0,0.6,0}
\definecolor{codegray}{rgb}{0.5,0.5,0.5}
\definecolor{codepurple}{rgb}{0.58,0,0.82}
\definecolor{backcolour}{rgb}{0.95,0.95,0.92}
\lstdefinestyle{mystyle}{
  backgroundcolor=\color{backcolour},   commentstyle=\color{codegreen},
  keywordstyle=\color{magenta},
  numberstyle=\tiny\color{codegray},
  stringstyle=\color{codepurple},
  basicstyle=\ttfamily\footnotesize,
  breakatwhitespace=false,         
  breaklines=true,                 
  captionpos=b,                    
  keepspaces=true,                 
  numbers=left,                    
  numbersep=5pt,                  
  showspaces=false,                
  showstringspaces=false,
  showtabs=false,                  
  tabsize=2
}
\usepackage{tikz,xcolor,hyperref}

\definecolor{lime}{HTML}{A6CE39}
\DeclareRobustCommand{\orcidicon}{%
	\begin{tikzpicture}
	\draw[lime, fill=lime] (0,0) 
	circle [radius=0.16] 
	node[white] {{\fontfamily{qag}\selectfont \tiny ID}};
	\draw[white, fill=white] (-0.0625,0.095) 
	circle [radius=0.007];
	\end{tikzpicture}
	\hspace{-2mm}
}

\foreach \x in {A, ..., Z}{%
	\expandafter\xdef\csname orcid\x\endcsname{\noexpand\href{https://orcid.org/\csname orcidauthor\x\endcsname}{\noexpand\orcidicon}}
}

\begin{document}

\title{Open Source Variational Quantum Eigensolver Extension of the Quantum Learning Machine (QLM) for Quantum Chemistry}

\author[1,2,3]{Mohammad Haidar*}
\author[3]{Marko J. Rančić}
\author[4]{Thomas Ayral}
\author[2,5]{Yvon Maday}
\author[1]{Jean-Philip Piquemal*}

\authormark{HAIDAR \textsc{et al}}

\address[1]{\orgdiv{Sorbonne Université}, \orgname{Laboratoire de Chimie Théorique(UMR-7616-CNRS)}, \orgaddress{\state{4 place Jussieu-75005 Paris}, \country{France}}}
\address[2]{\orgdiv{Sorbonne Université, CNRS, Université Paris Cité}, \orgname{Laboratoire Jacques Louis Lions (LJLL)}, \orgaddress{\state{4 place Jussieu-75005 Paris}, \country{France}}}
\address[3]{\orgdiv{TotalEnergies}, \orgname{Tour Coupole La D\'{e}fense}, \orgaddress{\state{2 Pl. Jean Millier, 92078 Paris}, \country{France}}}
\address[4]{\orgdiv{Atos Quantum Laboratory}, \orgaddress{\state{Les Clayes-sous-Bois}, \country{France}}}
\address[5]{\orgdiv{Institut Universitaire de France},\orgaddress{\state{Paris}, \country{France}}}

\corres{Mohammad Haidar, Sorbonne Université, Laboratoire de Chimie Théorique(UMR-7616-CNRS) and Laboratoire Jacques Louis Lions (UMR-7598-CNRS), 4 place Jussieu, 75005 Paris, France. \email{mohammad.haidar@upmc.fr};\\ Marko J. Ran\v{c}i\'{c}, TotalEnergies, Tour Coupole La D\'{e}fense, 2 Pl. Jean Millier, 92078 Paris, France. \email{marko.rancic@totalenergies.com}; Jean-Philip Piquemal, Sorbonne Université, Laboratoire de Chimie Théorique(UMR-7616-CNRS). \email{jean-philip.piquemal@sorbonne-universite.fr}}


\abstract[Abstract]{Quantum Chemistry (QC) is one of the most promising applications of Quantum Computing. However, present quantum processing units (QPUs) are still subject to large errors. Therefore, noisy intermediate-scale quantum (NISQ) hardware is limited in terms of qubit counts/circuit depths. Variational Quantum Eigensolver (VQE) algorithms can potentially overcome such issues. Here, we introduce the OpenVQE open-source QC package. It provides tools for using and developing chemically-inspired adaptive methods derived from Unitary Coupled Cluster (UCC). It facilitates the development and testing of VQE algorithms and is able to use the Atos Quantum Learning Machine (QLM), a general quantum programming framework enabling to write/optimize/simulate quantum computing programs. We present a specific, freely available QLM open-source module, myQLM-fermion. We review its key tools for facilitating QC computations (fermionic second quantization, fermion-spin transforms, etc.). OpenVQE largely extends the QLM’s QC capabilities by providing: (i) the functions to generate the different types of excitations beyond the commonly used UCCSD ansatz; (ii) a new Python implementation of the "adaptive derivative assembled pseudo-Trotter method" (ADAPT-VQE). Interoperability with other major quantum programming frameworks is ensured thanks to the myQLM-interop package, which allows users to build their own code and easily execute it on existing QPUs. The combined OpenVQE/myQLM-fermion libraries facilitate the implementation, testing and development of variational quantum algorithms, while offering access to large molecules as the noiseless Schrödinger-style dense simulator can reach up to 41 qubits for any circuit. Extensive benchmarks are provided for molecules associated to qubit counts ranging from 4 up to 24. We focus on reaching chemical accuracy, reducing the number of circuit gates and optimizing parameters and operators between "fixed-length" UCC and ADAPT-VQE ansätze.}

\jnlcitation{\cname{%
\author{M. Haidar}, 
\author{M. J. Rančić}, 
\author{T. Ayral}, 
\author{Y. Maday}, and 
\author{J-P. Piquemal}} (\cyear{2022}), 
\ctitle{Open Source Variational Quantum Eigensolver Extension of the
Quantum Learning Machine (QLM) for Quantum Chemistry}, \cjournal{}, \cvol{}.}
\maketitle
\section{\label{sec:intro}Introduction}
Solving the Schrödinger equation to obtain the many-electron wavefunction is the central problem of modern quantum chemistry \cite{aspuru2005simulated, bassman2021simulating}. In practice, it is possible to achieve full accuracy for systems that contain few electrons through methods like the Full Configuration Interaction (FCI) and the Full Coupled-Cluster (FCC), which include all the electron configurations.
However, the computational cost grows exponentially---through the dimension of the FCI and FCC wave functions---when the number of electrons increases. For example, the number of possible Slater determinants increases as a function of the number of electrons $n_e$ and of orbitals $n_{o}$\cite{lehtola2017cluster}:
\begin{equation}
\displaystyle
\mathrm{dim}{\cal H} \approx \left(\frac{ n_o!}{(n_o - n_e/2)!(n_e/2)!}\right)^2
 \label{eq:dimH}
\end{equation}

Attempts to reach full accuracy on large systems clearly faces the so-called “exponential wall” that limits the applicability of the most accurate methods to more complex chemical systems. So far, the largest calculations performed with classical supercomputers have only included tens of billions of determinants\cite{gan2006lowest} with 20 electrons and 20 orbitals, with hopes to solve problems of a size close to a trillion determinants (24 electrons, 24 orbitals) in a near future thanks to the advances of massively parallel supercomputer architectures \cite{vogiatzis2017pushing}.
Given such constraints, other classes of methods have to be used to approximate the ground state wavefunctions for larger many-electron systems. They include: (i) Density-Functional Theory (DFT), which relies on the use of a single Slater determinant, and has been shown to be extremely successful while failing to describe strongly correlated systems \cite{hohenberg1964inhomogeneous, kohn1965self, schutt2018machine};
(ii) post Hartree-Fock methodologies such as the truncated Coupled-Cluster (CC) and the  Configuration Interaction (CI) approaches that, even though being still operational beyond the single Slater determinant, cannot be applied to large-size molecules, due to their extreme computational requirements in terms of Slater determinants \cite{bartlett1989alternative, kutzelnigg1991error, harrison1991approximating, olsen1996full, peris1999perturbatively, taube2006new, bartlett2007coupled, schriber2016communication}. A good example is provided by the “gold standard” methods denoted coupled-cluster single-, double-, plus perturbative triple-excitations CCSD(T). Indeed, CCSD(T) is able to handle a few thousand basis functions, but at the cost of an enormous operation count that is limited by the large requirements in term of data storage\cite{nagy2019approaching}.
No matter the choice of chemical basis sets (STO-3G, 6-31G, cc-pVDZ, and beyond...), these methods are insufficient to reach accurate enough results for large molecules.
 
A paradigm shift
proposed by Feynman \cite{feynman2018simulating, benioff1980computer} is to use quantum computers for simulating quantum systems. 
This has prompted the community to use  quantum computers in order to solve quantum chemistry wavefunction problems. Intuitively, the advantage comes from the fact that quantum computers can process “exponentially” more information than classical computers\cite{reiher2017elucidating}. 
Recent reviews provide background material about strategies for developing quantum algorithms dedicated to quantum chemistry. These approaches include techniques such as Quantum Phase Estimation (QPE), Variational Quantum Eigensolvers (VQE) 
or Quantum Imaginary Time Evolution (QITE)
\cite{mcardle2020quantum, cao2019quantum, bauer2020quantum, yeter2021benchmarking}.
All methods generally include three key steps:
(i) transforming the fermionic Hamiltonian and wavefunction into a qubit representation;
(ii) constructing circuits with one and two-qubit quantum gates;
(iii) using the circuits to generate a relevant wavefunction and measure the expectation value of a given  Hamiltonian. 
Importantly, currently available quantum computers remain in the Noisy Intermediate-Scale Quantum (NISQ) era and are limited by two main resources: the number of qubits and the circuit depth (i.e. number of quantum gates) \cite{whitfield2011simulation, preskill2018quantum, elfving2020will, bharti2022noisy}.
Among all available strategies, the VQE family of techniques is a very promising algorithm that can be applied to NISQ hardware\cite{mcardle2020quantum, peruzzo2014variational, mcclean2016theory, mcardle2020quantum, cerezo2021variational}.
Indeed, it is better suited to such devices than other algorithms such as 
QPE, thanks to its more modest requirements in terms of quantum processor coherence times. It has already been successfully executed on real NISQ quantum computers based on various technologies, including superconducting qubits\cite{google2020hartree, kandala2017hardware}, photons\cite{peruzzo2014variational} and trapped ions\cite{hempel2018quantum}.
The performance of VQE  algorithms is known to be highly dependent on the type of wavefunction ansatz.
In practice, several types of ansätze can be used including:
unitary coupled-cluster (UCC) wavefunctions\cite{romero2018strategies, anand2021quantum}, 
hardware-efficient\cite{kandala2017hardware, gard2020efficient},
qubit coupled cluster (QCC)\cite{ryabinkin2018qubit},
deep multi-scale entanglement renormalization ansatz (DMERA)\cite{kim2017robust} and
Hamiltonian variational\cite{wecker2015progress}.
Furthermore, there are recent hybrid approaches such as adaptive ansätze\cite{grimsley2019adaptive} and the subspace expansion method\cite{mcclean2017hybrid}, which also utilize the basic VQE scheme to build the wavefunction of the system.


In this article, we provide a newly-designed software package named "OpenVQE" that extends the Atos Quantum Learning Machine ("QLM"). QLM is a quantum computing platform that allows to write hardware-agnostic quantum programs, compile them in compliance with hardware constraints, and either emulate them classically with realistic noise models or execute them on actual quantum hardware. QLM is a mix between commercial tools and non-commercial ones provided by the Atos company. OpenVQE is fully compatible with the freely available open source component of QLM called "myQLM", which is available online \cite{myQLM}.
myQLM is a quantum software stack for writing, simulating, optimizing, and executing quantum programs. It provides, through a Python interface:
(i) powerful semantics for manipulating quantum circuits, with support for universal as well as custom gate sets, abstract parameters, advanced linking options, etc;
(ii) a versatile execution stack for running quantum jobs, including an easy handling of observables, special plugins for carrying out NISQ-oriented variational methods (such as VQE or QPE), and an easy application programming interface (API) for writing customized plugins (e.g., for compilation or error mitigation), as well as for connecting to any Quantum Processing Unit (QPU);
(iii) a seamless interface to available quantum processors and major quantum programming frameworks\cite{QLMbinders}.
The capacity of QLM/myQLM to interoperate with other packages such as Qiskit\cite{wille2019ibm}, Cirq\cite{cirqql} and PyQuil\cite{smith2016practical} allows researchers from different horizons to communicate, interact, exchange, and work more efficiently and faster.
Furthermore, myQLM acts as a universal platform for users who want to access various quantum computers (IBM, Google etc...), providing  them simple tools to run directly their jobs.
myQLM comes with fermionic second quantization tools, which are helpful for solving quantum chemistry problems. These tools are now included in a new specialized open-source module dedicated to quantum chemistry called "myQLM-fermion". myQLM-fermion will be described in the present work; OpenVQE is designed to work synergistically with it.
myQLM-fermion includes the key QLM resources that are important for quantum chemistry developments.
More specifically, we use two important tools provided by myQLM-fermion which are useful for quantum resource reductions\cite{romero2018strategies}: 
(i) an active space (AS) selection approach that is useful for the reduction of the number of qubits;
(ii) an MP2 pre-screening approach that involves an implementation of second order Møller-Plesset perturbation (MP2) amplitudes as an input guess that is useful for faster parameter optimization.
Therefore, by building on top of  the initial myQLM blocks, we were able to design our OpenVQE package, which provides the community with advanced VQE tools dedicated to solve quantum chemistry problems.
OpenVQE especially focuses on giving access to modules enabling the use of the UCC family and to adaptive ansatz algorithms.
These modules include various features such as:
\begin{enumerate}
\item different types of UCC generators (truncated to single and double excitations):
(i) Unitary Coupled Cluster  Singles and Doubles (UCCSD \cite{nooijen2000can}) approaches;
(ii) Unitary Pair CC with Generalized Singles and Doubles Product (k-UpCCGSD\cite{lee2018generalized}) techniques; 
(iii) the possibility of including excitations that form spin-complemented pair interactions \cite{grimsley2019adaptive} and excitations that conserve only singlet spin symmetry of UCCGSD (see appendix of \cite{tang2021qubit});
(iv) Qubit Unitary Coupled Cluster Singles and Doubles QUCCSD \cite{Xia};
\item adaptive VQE algorithms (namely Adaptive Derivative-Assembled 
Pseudo-Trotter-VQE (ADAPT-VQE)), with different operator pools including:
(i) Unitary fermionic operators (fermionic-ADAPT-VQE\cite{grimsley2019adaptive});
(ii) Anti-Hermitian Pauli operators (qubit-ADAPT-VQE\cite{tang2021qubit,shkolnikov2021avoiding}).
\end{enumerate}
These modules are structured in simple Python classes that require only to write a few lines of code and provide therefore examples enabling a rapid implementation of additional algorithms on the basis of myQLM basic tools. 

The goal of this paper is to showcase the novel OpenVQE/myQLM-fermion  combined quantum simulator package that facilitates the test, development and measurement of the computational requirements of the different VQE flavors towards their future potential implementation on real supercomputers. In particular, it allows to compare the accuracy of the total molecular energies obtained thanks to various variants of the UCC method with popular quantum chemistry methods on classical computers.
The GitHub websites of OpenVQE\cite{openvqe} and myQLM-Fermion\cite{myqlm-fermion} are public, under "MIT" and "Apache-2.0" licenses, respectively. 

The paper is organized as follows: 
we review the theory of UCC ansätze including the different types of excitations, the ADAPT-VQE approach and the different steps needed to implement it (section \ref{sec:review}).
We then introduce the QLM library, specifically its myQLM-fermion open source extension, with a focus on its quantum chemistry tools, which we use develop OpenVQE (see section \ref{sec:qlm}).  
Then, based on myQLM-fermion components, we describe the OpenVQE package, with its implementation of advanced UCC ansätze (static or adaptive) (section\ref{sec:OpenVQE}).
Finally, executing the OpenVQE library on our in-house HPC architecture (a Quantum Learning Machine at TotalEnergies), we simulate a full set of molecules ranging from 4 to 24 qubits. 
We first show the properties of the simulator applied to a set of molecules.
Second, we describe how OpenVQE/myQLM-fermion can use active space selection and MP2 pre-screening initial guesses for different test molecules.
Third, using our ADAPT-VQE module, we compare the fermionic and qubit-ADAPT-VQE results obtained on a set of  molecules in terms of chemical accuracy, number of variational parameters, operators and quantum gates.
Finally, we compare the group of "fixed-length" ansätze (using UCC modules) to the ADAPT-VQE method (section\ref{sec:adaptvqe}). 
We conclude with a summary table comparing the features of the OpenVQE/myQLM-fermion packages with those implemented in other software packages, and we give some perspectives towards the development of new types of UCC ansätze and/or new variational algorithms within OpenVQE. 

\section{Review: Unitary Coupled Cluster (UCC) and Adaptive Derivative-Assembled Pseudo-Trotter (ADAPT) within the VQE Algorithm}\label{sec:review}

We will use the  following notations in the article:
the indices \textit{i} and \textit{j} refer to the occupied spin-orbitals in the Hartree-Fock (HF) state; indices \textit{a} and \textit{b} refer to the unoccupied (or virtual) spin-orbitals in the HF state. 
Indices \textit{p,q,r} and \textit{s} are always related to spin-orbitals unless mentioned otherwise. When the orbitals are associated to an $\alpha$-like symbol ($p_{\alpha}$, $q_{\alpha}$….) they refer to spin-up electrons, and those with a $\beta$-like ($p_{\beta}$, $q_{\beta}$….)  refer to spin-down electrons.  $n_e$ and $n_o$ (as given in the introduction) represent the number of electrons and orbitals, respectively.
We also denote by  $\eta$ and $N_A$ the number of active electrons and number of active spin-orbitals, respectively,  and  by $n$ the number of qubits.
In qubit representation, $|\psi_\mathrm{ref}\rangle$ denotes the $n$-qubit initial state. It is usually a computational basis state $|q_0, q_1, q_2, q_3, ...q_n\rangle$. 
We call $|\psi(\theta)\rangle$ the trial wave function and  $\theta$ its variational parameter, while  $U(\theta)$ denotes a parameterized unitary operator. 

\subsection{The Variational Quantum Eigensolver (VQE) with the Unitary Coupled Cluster (UCC) Ansatz} \label{subsec:VQE}

\subsubsection{VQE-UCC in a Nutshell}

The Variational Quantum Eigensolver (or VQE) has been first developed by Peruzzo et al. \cite{peruzzo2014variational}
This hybrid quantum classical algorithm  aims at finding the ground state energy $E_0$ of a given Hamiltonian $H$. 
In the context of quantum chemistry, the Hamiltonian is often written in the following fermionic second-quantized form:  
\begin{equation}
\label{eq:Ham1}
H = \sum_{pq} h_{pq} c^{\dagger}_pc_q + \frac{1}{2}\sum_{pqrs} h_{pqrs}c_p^{\dagger}c_q^{\dagger}c_r c_s,
\end{equation}
where $c_p^{\dagger}$ ($c_q$) are anti-commuting operators that create (annihilate) electrons in spin-orbital $p(q)$, respectively.
The symbols $h_{pq}$ and $h_{pqrs}$ denote the one- and two-body integrals of the corresponding operators and spin-orbitals in Dirac notation, respectively.
These integrals can be easily computed on a classical computer.

The VQE method is based on quantum variational theory, whose starting point is the Rayleigh-Ritz variational principle, which states that 
\begin{equation}
\label{energyoptimized}
    \langle \psi(\theta)| H |\psi(\theta)\rangle \ge E_0,
\end{equation}
namely the energy of the normalized “trial” or "ansatz" wave function $ |\psi(\theta)\rangle$ provides an upper bound for the ground state energy $E_0$.
Variational methods leverage this principle through an iterative process that minimizes the expectation value 
$ E(\theta)=\langle\psi(\theta)| H |\psi(\theta)\rangle $ with respect to the parameterized trial state $ |\psi(\theta)\rangle$.
This process is supposed to converge to an optimal parameter set $\theta^{*}$ and hence give an approximation $E(\theta^{*})$ of $E_0$. 

One of the core challenges of VQE is the choice of the ansatz $|\psi(\theta)\rangle$.

One popular ansatz is the unitary coupled cluster (UCC) approach.
It is inspired from the classical coupled cluster (CC) computational method \cite{Cizek1966,Cizek1980,bartlett1989alternative,bartlett2007coupled}, which has been shown to be a good choice for reaching accurate results.
The unitary version of CC, UCC, was originally introduced in the field of quantum chemistry computations \cite{taube2006new}.
It was later brought to the field of quantum computing  \cite{Yung2014,peruzzo2014variational} because of its unitary character (as opposed to the CC method).
A comprehensive review of the UCC method is given in \cite{romero2018strategies,anand2021quantum}.
UCC yields variational states prepared by using unitary evolution under a sum of fermionic terms representing excitations of the initial Hartree-Fock (HF) state.
The UCC wavefunction can be expressed as follows:
\begin{equation}
\label{eqexpT}
\displaystyle
| \psi_\mathrm{UCC} \rangle = e^{T-T^\dagger} | \psi_\mathrm{HF}\rangle,
\end{equation}
where $T$ is the so-called cluster operator and $T^{\dagger}$ is its Hermitian conjugate.
As $T$ is  the  sum of the excitations at different levels
$T = T_1$ + $T_2$ + \dots +$T_x$ until $x$ excitations, then 
\begin{equation}
\label{nexcitations}
 T-T^{\dagger} = \sum_{a,i}^{n_{o}-1} \theta_{ai} (c_a^\dagger c_i - c_i^\dagger c_a) +  \sum_{a,b,i,j}^{n_{o}-1} \theta_{abij} (c_a^\dagger c_b ^\dagger c_i c_j - c_j^\dagger c_i^\dagger c_b c_a) + \dots
\end{equation}
where $a,b \in virt, i,j \in occ$, $\theta_{ai}$  and $\theta_{abij}$ are variational parameters.

This form of variational state ensures that UCC applied within VQE can systematically approximate the true ground eigenstate of the system. This is not guaranteed in the case for example of the hardware efficient ansätze \cite{kandala2017hardware,barkoutsos2018quantum,ganzhorn2019gate,sennane2022calculating}, whose form is not inspired by chemistry considerations.
Moreover, the hardware-efficient ansatz includes many optimization parameters, and can therefore suffer from the “barren plateau”\cite{mcclean2018barren,cerezo2021cost} problem.
The UCC optimization is not straightforward on a classical computer even for low-order cluster operators, due to the non-truncation of the Baker–Campbell–Hausdorff (BCH) series \cite{Taube2006}. 
On the other hand, a UCC state can be easily prepared on a quantum device, with first experimental demonstrations on a trapped-ion quantum simulator\cite{Yung2014,shen2017quantum}.
There, UCC-VQE was shown to be able to reach a final state with a good overlap  $|\langle\psi_{UCC} | \psi_0 \rangle|$ between the converged UCC solution and the FCI solution $|\psi_0\rangle$ (see for example Figure 3 in \cite{Yung2014}). \\\\

The implementation of the UCC wavefunction on a quantum computer is an important step in the VQE process. Further implementation aspects are described in section \ref{sec:qlm}. In particular, Figure \ref{fig:QLM} summarizes all the VQE process.

The UCC wavefunction in Eq. (\ref{eqexpT}) is implemented on a quantum device by first constructing a circuit corresponding to the qubit representation of the HF state, by using (for instance) the Jordan Wigner (JW) representation\cite{fradkin1989jordan}, $\displaystyle| \psi_{HF}\rangle_Q = |11110000...\rangle$
where $|0/1\rangle$ refers to a single qubit being in the state 0 or 1.
In  the multi-qubit register, we assign each qubit to a spin-orbital, with orbital energies ranging from low to high.
The $|0\rangle$ state refers to an unoccupied spin-orbital, while $|1\rangle$ refers to an occupied spin-orbital. The JW transformation relates spin-1/2 operators to fermionic creation and annihilation operators. We will use JW throughout the article.
After having prepared $|\psi_{HF}\rangle_Q$ on the device, the unitary operator $U(\theta)= e^{T-T^\dagger}$ has to be implemented as well. This is done following these steps: $U(\theta)$ must be decomposed into operations that can be implemented as circuits compatible with present available quantum computers.
To do this, in practice,  $U(\theta)$ is approximated using the Trotter decomposition \cite{Evangelista2019,Grimsleytrotter,hatano2005finding}, which breaks up the exponential of a sum as an ordered product of individual exponentials. Such transformation provides a reasonable approximation to the full unitary operator that is disassembled into local unitary operators.
The formula of the trotterized   $U(\theta)$ with  $T(\theta)-T(\theta)^\dagger$ (Eq. \eqref{nexcitations})  is:
\begin{equation} 
\label{tettor}
U(\theta) =
\bigg(\prod_{\rho} e^{\frac{\theta_{\rho}}{t} (T_\rho-T_\rho^\dagger)}\bigg)^t  + \mathcal{O}\,(\frac{1}{t})
\end{equation} 
where $t$ is the number of  Trotter steps and $\rho$ corresponds to the elements of excitations introduced in Eq. \ref{nexcitations}.
Using $t > 1$ can, on paper, decrease the Trotterization error. Yet, for the current available NISQ devices, it is not recommended since it would involve additional gates induced by the UCC ansatz translation into a quantum circuit \cite{barkoutsos2018quantum,anand2021quantum}.
For instance, in reference \cite{barkoutsos2018quantum} (Figure 5), increasing the number of Trotter steps $t$ using a UCCSD ansatz for the H$_2$ molecule does not improve the convergence and the UCCSD energy. 
In the present work, we use a UCC ansatz truncated at the single and double excitation levels with a single Trotter step.
Once these approximations are performed, each exponential of a fermionic excitation operator (see Eq. (\ref{tettor})) can be directly implemented as a sequence of gates.
This is done using the standard "CNOT staircase method"  \cite{mcardle2020quantum,whitfield2011simulation,hempel2018quantum} or a more advanced method \cite{Yordan2019,Yordan2020, Yordan2021} to reduce the CNOT counts. 
Further details about the implementation of the VQE-UCC method will be given in subsection \ref{myqlm-fermion}.

\subsubsection{Unitary Coupled Cluster Approaches Truncated at Single and Double Excitations Levels}
\label{UCCsection}
In this subsection, we list the different versions of UCC that are truncated at the single and double excitation levels, by defining the excitation generator in each UCC version. We also review their recent use within variational algorithms.
We first focus on the UCCSD and QUCCSD ansätze to compare their gate complexity and CNOT counts.
We then focus on a second group of methods that includes UCCGSD and k-UpUCCGSD which, as shown in some reviews, perform better than the first group of ansätze in terms of chemical accuracy. However, such approaches require an increased circuit depth compared to the first group of techniques.
These features have motivated us to implement these versions of UCC together with ADAPT-VQE (presented in the next section) in our OpenVQE package  (see sec \ref{sec:OpenVQE}). 

Unitary Coupled Cluster Single and Double was the first commonly used ansatz by the quantum chemistry community for quantum computing, and it was successfully tested experimentally on quantum computers using VQE \cite{Yung2014,hempel2018quantum}.
Chemically, the UCCSD ansatz includes few excitations of fermionic single and double operators, which occur only from occupied to unoccupied (virtual) spin-orbitals on the top of the HF state as given in Eq.~\eqref{nexcitations}. By assuming a single-step Trotter approximation (Eq. (\ref{tettor})), UCCSD is given by the product of single and double fermionic evolutions
\begin{equation}
\label{uccsdtrotter}
U(\mathbf{\theta}) = \prod_{a,i} e^{\big(\theta_{ai} (c_a^\dagger c_i - c_i ^\dagger c_a)\big)}\prod_{a,b,i,j}  e^{\big(\theta_{abij} (c_a ^\dagger c_b^\dagger c_ic_j - c_j^\dagger c_i^\dagger c_b c_a)\big)}. 
\end{equation}
The Qubit Unitary Coupled Cluster Single and Double method is a simplified version of the UCCSD ansatz (Eq. \eqref{uccsdtrotter}) in which the single and double excitations $q^\dagger_a$ ($q_i$) used to construct the trial wavefunction
 \begin{equation}
\label{quccsdtrotter}
U(\mathbf{\theta}) = 
\prod_{a,i} e^{\big(\theta_{ai} (q_a^\dagger q_i - q_i ^\dagger q_a)\big)}\prod_{a,b,i,j}  e^{\big(\theta_{abij} (q_a ^\dagger q_b^\dagger q_iq_j - q_j^\dagger q_i^\dagger q_b q_a)\big)},
\end{equation}
 are the   creation (annihilation) operators without the inclusion of Pauli-$Z$ terms after JW transformation (see their expressions in chapter two, section 2.2 (Eqs. 2.11 and 2.12) of \cite{Yordan}). These excitations can be achieved by particle-preserving exchange rotation gates in the qubit space. 
The actual role of Pauli-$Z$ is to conserve the parity from fermionic creation (annihilation) $c^\dagger$ ($c$) operators, while preserving the time and particle symmetries. 
By comparing QUCCSD to UCCSD ansatz VQE performance for a few simple molecules, it has been shown  \cite{Xia} that the two ansätze could achieve a nearly identical accuracy in the estimation of the ground state energies.
Even though the parity property is missed in the qubit evolution excitations, it was found, as is
described in Figures 1 and 2 of \cite{Xia}, that the advantage of using QUCCSD is to reduce the CNOT counts with respect to the CNOT staircase circuit to implement fermionic evolutions.
This demonstrates that QUCCSD is sufficient to approximate the exact FCI wavefunction  with a smaller number of gates in the  circuit and it might therefore be more favorable to use on current NISQ devices. 
However it has been shown that the operators in UCCSD and QUCCSD ansätze may not be ordered in practice and could decrease the precision in energy. Attempts for keeping ordered operators have been recently proposed. \cite{romero2018strategies,ryabinkin2018qubit,grimsley2019adaptive,William2020}.
Moreover it has been shown that an unnecessarily large number of parameters and operators occur when these ansätze were used, especially when system size gets larger. This is because the information about
the specific chemical system such as point group symmetry were not taken into account.
For example, as is shown in Table I of \cite{cao2022progress}, when the point group symmetry is used in UCCSD, the number of parameters reduce by at least 20\%, which accelerates the optimization procedure and even brings little improvement, compared to UCCSD energies calculated without parameter reduction. 

The Unitary Coupled Cluster Generalized Single and Double (UCCGSD) ansatz state has been first mentioned in Reference~\cite{nooijen2000can}, then later in Reference~\cite{lee2018generalized} for using it in quantum computing applications.
It consists into excitations that occur from occupied to occupied, and occupied to unoccupied levels, as is expressed in Equation\ref{uccsdtrotter}, or from unoccupied to unoccupied levels:
\begin{equation}
\label{uccgsd}
    \sum_{pq} \theta_{qp} (c^\dagger_q c_p-c^\dagger_p c_q) +  \sum_{pqrs} \theta_{rspq} (c_r ^\dagger c_s ^\dagger c_q c_p - c_p ^\dagger c_q ^\dagger c_s c_r).
\end{equation}    
$k$-UpCCGSD stands for Unitary Pair Coupled Cluster with Generalized Single and Double Product  \cite{lee2018generalized}, where $k$ denotes the products of unitary paired generalized double excitations, along with the full set of generalized single excitations.
In other words, in $k$-UpCCGSD, single excitation operators are fully generalized without any specific constraint on the choice of occupied and virtual orbitals, but the double excitation operators are generalized and restricted to transitions of pair electrons acting on the same orbital:
\begin{equation}
\label{kupccgsd}
T_2 -T_2^\dagger = \\
\sum_{q_\alpha q_\beta p_\alpha p_\beta}\theta_{q_\alpha q_\beta p_\alpha p_\beta} (c_{q_{\alpha}} ^\dagger c_{q_{\beta}} ^\dagger c_{p_{\beta}} c_{p_{\alpha}} - c_{p_{\alpha}} ^\dagger c_{p_{\beta}} ^\dagger c_{q_{\beta}} c_{q_{\alpha}})
\end{equation}
It was recently demonstrated \cite{lee2018generalized,Greene} that the UCCGSD  ansatz leads to more accurate results for ground state energies than the simpler UCCSD. In the same work, it was shown that starting from $k=3$ (i.e 3-UpCCGSD) brings better accuracy than the UCCSD ansatz (see for example table 4 for H$_4$ molecule using 6-31G basis set).
Indeed, such approaches were also engineered not only for ground states but also to tackle excited states.
They have been used in the Variational Quantum Deflation algorithm (VQD), which is used to determine the energies of the ground and excited states \cite{Chan}.

\subsection{Adaptive Derivative-Assembled Pseudo-Trotter-VQE (ADAPT-VQE)}
\label{ADaPT-VQE_review}
Instead of a fixed-length ansatz, i.e. where one chooses UCCSD, UCCSDT or UCCSDTQ anszatz (length equal 2, 3 or 4 respectively), variational algorithms with a variable-length ansatz have been proposed by the community\cite{Arthur2019,Ilya2020,Robert2020,SukinSim,JieLiu} 
There are two reasons for that:
(1) the fixed-length ansätze use unnecessary excitations that can be considered as redundant terms as they do not contribute to a better approximation of the FCI wavefunction.  This creates longer circuits and consumes a greater number of variational parameters;
(2) fixed-length ansätze such as UCCSD or QUCCSD cannot provide a good chemical accuracy for strongly correlated systems\cite{Xia} (i.e. see for example the H$_6$  molecule in section \ref{sec:adaptvqe}). 
This problem occurs at long bond lengths and could be solved by including higher-order excitations and/or using multiple-step Trotterization in the UCCSD or QUCCSD ansatz.
Yet, both suggestions are problematic since they would necessarily deepen the  circuits and increase the number of variational parameters. 
Fixed-length ansätze like 3-UpCCGSD (or k$>$3) and UCCGSD might also provide very good accuracy at these bond lengths, because the excitations considered are general. However, this yields  many redundant terms during optimization, and as the system size increases, the number of operators and parameters, as well as the circuit depth, increase.  
The ADAPT-VQE approach is an important step toward solving these issues \cite{grimsley2019adaptive,tang2021qubit}.

The ADAPT-VQE algorithm constructs the molecular system's wavefunction dynamically and can in principle avoid redundant terms. It is grown iteratively in the form of a disentangled UCC ansatz  as  given in Eq. (\ref{FCIADAPT}). At each step, an operator or a few operators are chosen from a pool: the operator(s) contributing to the largest energy deviations is (are) chosen and added gradually to the ansatz until the exact FCI wavefunction has been reached 
\begin{equation}
\label{FCIADAPT}
\prod_{k=1} ^{\infty}\prod_{pq}\left(e^{\theta_{pq}(k)\hat{A}_{p,q}}\prod_{rs}e^{\theta_{pqrs}(k)\hat{A}_{pq,rs}}\right) |\psi_\mathrm{HF}\rangle,
\end{equation}
which is in the form of a long product of one-$\hat{A}_{p,q}$ and two-body $\hat{A}_{pq,rs}$ operators generated from the pool of excitations, where each of the variational parameters $\{\theta_{pq},\theta_{pqrs}\}$ is associated to an operator. 

Let us describe briefly the procedure to implement the ADAPT-VQE (see details in \cite{grimsley2019adaptive}).
A reference Hartree-Fock $\displaystyle |\psi_{HF} \rangle$ state needs to be chosen first in the qubit representation that preserves the number of electrons of the system, which is an essential component for implementing the algorithm.
At this stage, it is then possible to construct the ansatz by starting the ADAPT-VQE-iteration at $k=1$.
The circuit is initialized by the identity $\displaystyle U^{(0)}(\theta) = I$, corresponding to the HF state as an initial state.
Then to start constructing the ADAPT wavefunction, the algorithm  starts (by a "while loop" until exit):\\

1) Then the trial state with the current ansatz on the quantum simulator $ \displaystyle |\psi^{(k-1)}\rangle = U^{k-1} \left(\theta_{k-1}\right)| \psi_{HF} \rangle$, where $\displaystyle \theta_{k-1}$ comes from the previous VQE iteration, has to be prepared;\\

2) In order to obtain the gradient (i.e  the energy derivative with respect to $\displaystyle \theta_{k-1}$), one measures the commutator  (in the qubit representation) between the Hamiltonian $H$ (Eq.~\ref{Pauli-Hamiltonian}) and each of the operators in the pool, of size $m$, which is defined by $\displaystyle A_{m}= \{A_{m}(p,q),A_{m}(p,q,r,s)\}$.
The energy gradient formula reads 
\begin{equation}
\displaystyle
\frac{\partial E^{(k-1)}}{\partial \theta_{m}}  = \langle \psi\left(\theta_{k-1}\right)| [H,A_m]|\psi\left(\theta_{k-1}\right)\rangle .
\end{equation}\\

3) If the norm of the gradient vector 
$||g^{(k-1)}|| = \sqrt{\left(\frac{\partial E^{(k-1)}}{\partial \theta_{1}}\right)^2+ \dots + \left(\frac{\partial E^{(m)}}{\partial \theta_{m}}\right)^2}$
becomes smaller than a threshold, $\epsilon$, then the algorithm exits the loop.
If not, the operator with the largest gradient 
$ \max_{\theta_i}\left(\big|\frac{\partial E^{(1)}}{\partial \theta_1}\big|,\dots, \big|\frac{\partial E^{(k-1)}}{\partial \theta_m}\big|\right) $ is selected,
and added to the left end of the ansatz, with a new variational parameter $\displaystyle\theta_k =\theta_i$. The wavefunction corresponding to ADAPT-VQE is given by
\begin{equation}
\displaystyle
|\psi^{(k)}\rangle = e^{\theta_k A_K}|\psi^{(k-1)}\rangle= e^{\theta_k A_k}\cdots e^{\theta_3 A_3} e^{\theta_2 A_2}e^{\theta_1 A_1} | \psi_{HF} \rangle.
\end{equation}

4) Perform a VQE experiment to re-optimize all parameters $\{\theta_k, \theta_{k-1},\dots,\theta_{2},\theta_{1}\}$ in
the ansatz and when this is over we go back to step 1.

Historically, fermionic-ADAPT-VQE \cite{grimsley2019adaptive} was the first algorithm in the family of adaptive ansätze.
The name “fermionic” comes from the excitation pool operators formed as spin-complement pairs of single and double fermionic evolutions (see Eqs. S1 and S2 are given in the Supplementary Material\cite{supplementary}). The fermionic-ADAPT-VQE has been shown to bring accurate chemical results for several molecules such as for H$_4$, LIH and H$_6$ molecules (as shown in figure 2 of \cite{grimsley2019adaptive}) by using an ansatz  constituted of fewer parameters and shorter circuit depth than for the corresponding UCCSD results.

After this fermionic version, qubit-ADAPT-VQE\cite{tang2021qubit,shkolnikov2021avoiding} was introduced to reduce the number of gates compared to fermionic-ADAPT-VQE. 
In qubit-ADAPT-VQE, the  pool used is a collection of anti-Hermitian operators where each operator is a tensor product of Pauli matrices. Each such operator is transpiled into a defined number of CNOT gates using the CNOT staircase method, but unlike the fermionic-ADAPT-VQE, the qubit-ADAPT-VQE produces a smaller number of gates at the end of the process.
Qubit-ADAPT-VQE has been tested for several molecules \cite{tang2021qubit} and lowers the gate counts by one order of magnitude compared to fermionic-ADAPT-VQE. 
However, qubit-ADAPT-VQE requires a higher number of parameters and iterations in order to reach a given level of accuracy.

The cost of fermionic- and qubit-ADAPT-VQE is the number of measurements needed to compute the energy gradients. It scales as $O(n^8)$ \cite{grimsley2019adaptive, tang2021qubit} (where $n$ represents the number of qubits).
The choice of pool in qubit-ADAPT-VQE has also been important in the research to reach and approximate the FCI wavefunction, with a view to not only reduce the circuit depth  compared to fermionic-ADAPT-VQE  but also to minimize the number of operators in the pool without losing the completeness properties of operators. 
Reducing the pool size from $\mathcal{O}(n^4)$ into 2$n-2$ operators and maintaining the necessary and sufficient conditions for completeness has been achieved in a recent work \cite{shkolnikov2021avoiding}.
This new  pool size reduces the number of measurements from $O(n^8)$ to $O(n^5)$ for a given molecular system.
In the same reference, it was shown that  incorporating symmetries into the pool solves the convergence problems caused by the lack of symmetries.

\section{Algorithms for Quantum Chemistry on the Atos Quantum Learning Machine}\label{sec:qlm}
In this section, we briefly review the tools of the Atos Quantum Learning Machine (QLM) that are relevant to quantum chemistry computations. We will then describe in the next section (sec. \ref{sec:OpenVQE}) how we built an advanced chemistry module upon these tools.

\subsection{A Quantum Programming Environment and a Powerful Simulator}

QLM is a complete environment designed for quantum software programmers, engineers and researchers \cite{atosqlm}. It includes a wide variety of low-level tools useful for writing, compiling and optimizing quantum circuits \cite{Martiel2020a, Brugiere2021, Vandaele2021, DeBrugiere2022, DeBrugiere2022a}. These tools come in the form of so-called "plugins" that can be combined or stacked upon another to construct a quantum compilation chain.

\begin{figure*}[h!]
\centering
\includegraphics[width=1.0\textwidth]{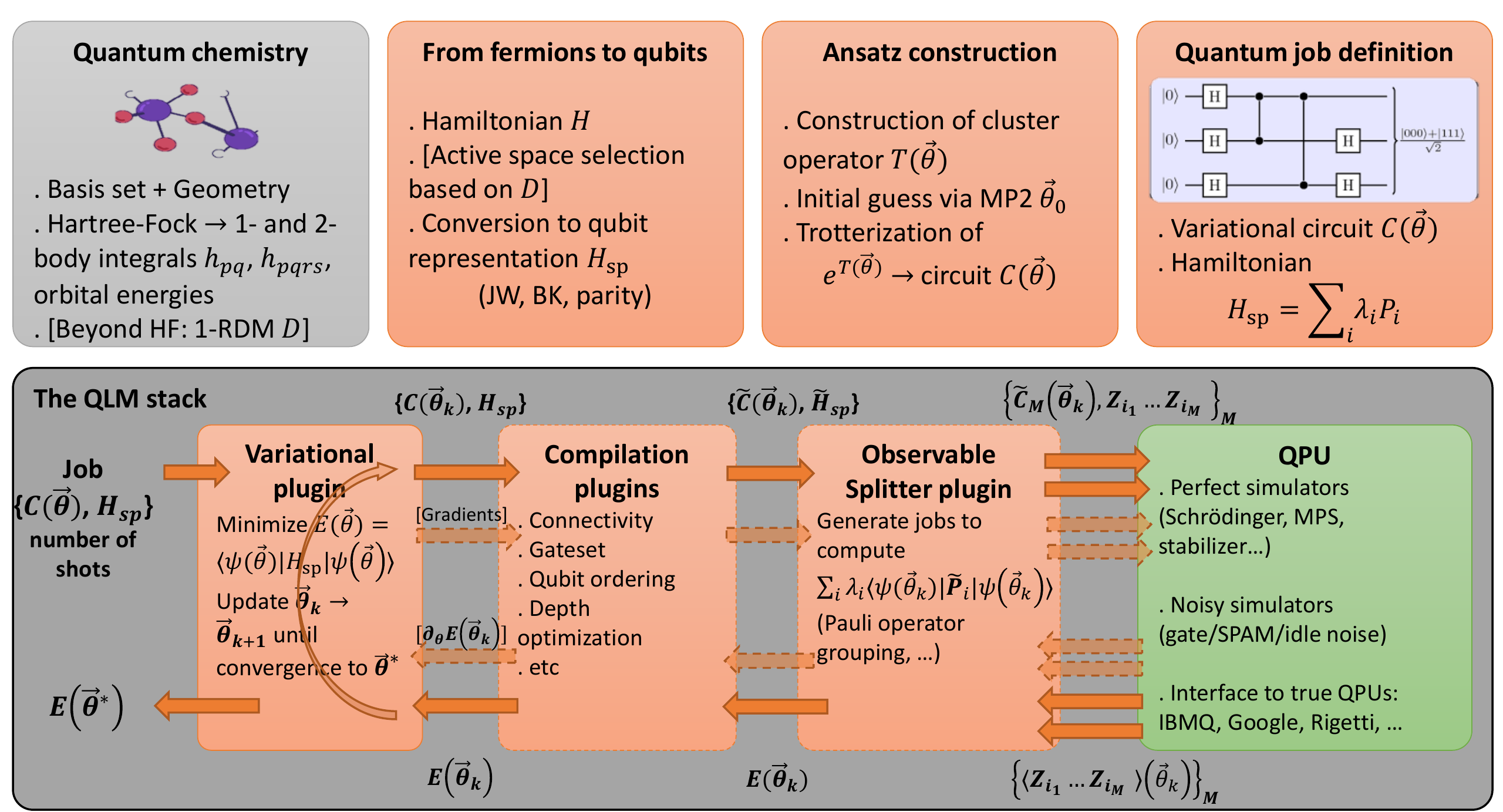}
\caption{The QLM workflow for quantum chemistry. \emph{Top row}: Steps to prepare a variational quantum job containing a parameterized circuit and the Hamiltonian whose ground state energy one wants to approximate. The leftmost (grey) box uses standard third-party quantum chemistry modules. Orange boxes stand for QLM libraries. \emph{Bottom row}: QLM stack, with plugins (orange boxes) that pre- and post-process the job and results, and a QPU (green box) that executes the quantum job and returns a result.}
\label{fig:QLM}
\end{figure*}
QLM can simulate quantum circuits in a noiseless or noisy fashion.
The noiseless simulators come in different flavors, with "Schrödinger style" simulators (storing the wavefunction either in a dense fashion or using Matrix Product States \cite{Vidal2003}), a "Feynman-style" simulator \cite{Rudiak-Gould2006}, a binary decision diagram simulator \cite{Miller2006}, a Clifford simulator, etc. The Schrödinger-style dense simulator can reach up to 41 qubits for any circuit, while the other simulators can reach much larger qubit counts depending on the circuit properties (entanglement, gateset, etc).
The noisy simulators enable the emulation of realistic quantum noise, a crucial tool in the current NISQ era. Gate noise, State Preparation and Measurement (SPAM) noise and idling noise can be taken into account via density-matrix (resp. stochastic) simulations that can handle circuit with up to 20 (resp. 40) qubits.
Most importantly, the interface of the simulators (also called "Quantum Processing Units" or QPUs) is such that they can easily be swapped for actual experimental QPUs without modifying the quantum program. QLM provides an interface to various hardware processors including those constructed by IBM, Rigetti etc... Its circuits are also interoperable\cite{QLMbinders} with other quantum circuit descriptions like Qiskit \cite{wille2019ibm}, Cirq \cite{cirqql} and PyQuil \cite{smith2016practical}, as described in more detail in section SI in the Supplementary Material\cite{supplementary}.
Moreover, QLM comes with quantum application libraries that enable the easy exploration of potential use cases of quantum computing. These libraries help generate quantum programs in fields ranging from combinatorial optimization to quantum chemistry and condensed-matter physics. 
A summary of the QLM workflow for quantum chemistry is provided in Figure  \ref{fig:QLM}. The top row represents the steps one has to go through to handle a quantum chemistry problem using \verb+myQLM-fermion+ (bracketed terms represent optional steps), including the transformation of the electronic structure Hamiltonian (Eq.~\ref{eq:Ham1}) to a spin representation:
\begin{equation}
\label{Pauli-Hamiltonian}
H_\mathrm{sp} = \sum_j \lambda_j P_j,
\end{equation}
with $P_j=\prod_{i=0}^{n_o-1}\sigma_i^j$, and $\sigma_i$ a Pauli matrix applied on qubit $i$.
At the end of these steps, one obtains a "job" comprising a parameterized quantum circuit (that implements a UCC ansatz) $C(\vec{\theta})$ and a Hamiltonian $H_\mathrm{sp}$ represented as a sum of Pauli terms. This job can then be fed to the QLM stack.

This stack, represented in the bottom row, can be constructed by the user at will depending on the QPU they want to execute the job on. A minimal stack consists of a QPU (without plugins), which can handle minimal jobs containing native quantum circuits (with gates native to the hardware and compliant with its connectivity) and $Z$-axis observables. To handle more sophisticated jobs, one can stack "plugins" on top of the QPU.
Each plugin has a well-defined role, such as compiling the circuit to a given gateset, rewriting it to comply with a given connectivity (see "Compilation plugins"), or generating all the elementary jobs required to compute the expectation value of a general observable (see "Observable Splitter plugin"). 
Finally, some plugins can handle variational jobs corresponding to the VQE algorithm. Such "variational plugins" handle the update of the variational parameters based on the result of the previous steps. They can optionally generate jobs to compute the gradient of a given expectation value if gradient-based optimization is asked for.

\begin{figure*}[h!]
\centering
\includegraphics[width=0.8\textwidth]{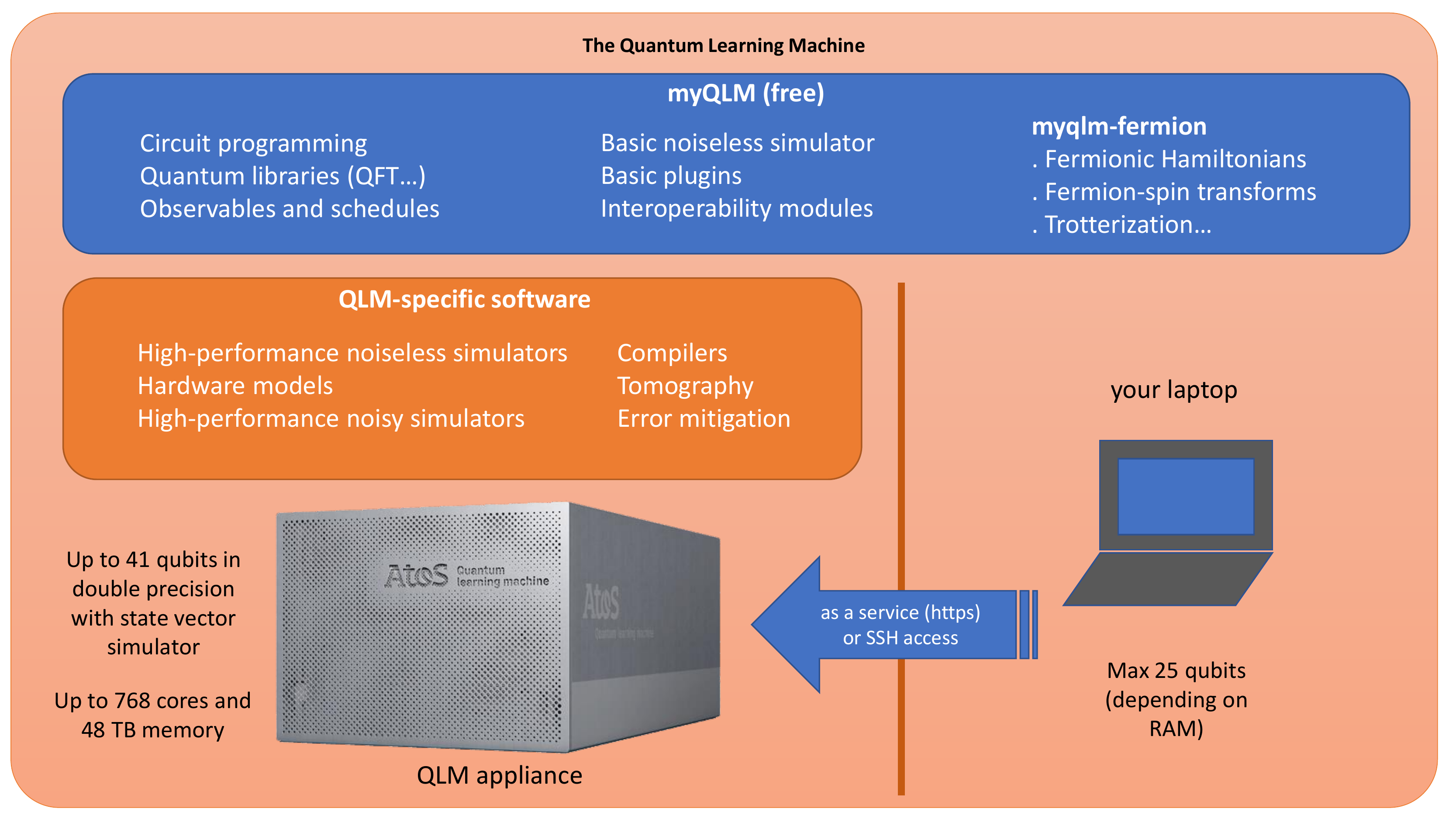}
\caption{Overview of the QLM environment. Documentation of my-QLM/myQLM-fermion is given in \cite{myQLM,myqlm-fermion}.}
\label{fig:QLM_myQLM}
\end{figure*}

In practice, the QLM environment is composed of software modules and a powerful classical HPC hardware appliance, as described in figure ~\ref{fig:QLM_myQLM}.
Among the software modules, some are available and downloadable for free as Python packages. They come under the name of \verb+myQLM+ \cite{myQLM}. \verb+myQLM+ includes modules for programming quantum circuits, with quantum libraries (for performing e.g quantum arithmetic operations, etc), for describing quantum observables and analog quantum schedules. It also comes with a basic noiseless simulator for quantum circuits, basic plugins (e.g for splitting observables depending on the commutation relations of the Pauli operators), and modules for interacting with other quantum frameworks like Qiskit, Cirq, etc.
Finally, \verb+myQLM+ contains the aforementioned \verb+myQLM-fermion+ module.

\verb+myQLM+ can be installed on a laptop or on a local cluster. The size of the circuits that can be simulated with the QPU simulator of \verb+myQLM+ is limited by the available Random Access Memory (RAM) (for a standard laptop, this corresponds to about 25 qubits). 
The Quantum Learning Machine is also a powerful HPC appliance. It comes with extra software in addition to \verb+myQLM+: advanced noiseless simulators, modules for describing hardware models, noisy simulators (both gate-based and analog), advanced compilers, tomography and error mitigation modules, among others.
The appliance itself can simulate circuits with up to 41 qubits when using state vector simulators, and more when using more specific representations of the wave function (like Matrix Product States or stabilizers).
The software module we present below, \verb+OpenVQE+, relies only on \verb+myQLM+ modules, which means it can be used by simply downloading \verb+myQLM+ on one's personal machine. Therefore, the results shown for small qubit counts can be (up to the run times) reproduced on a personal laptop.
In practice, to execute the simulations, we used the QLM advanced noiseless statevector simulator so as to reach large qubit counts and gain speed.

\subsection{myQLM-fermion: An Open Source QLM module for Fermionic Many-body Problems}
\label{myqlm-fermion}
The tools for quantum chemistry are collected into an open-source module called "myQLM-fermion", which is part of myQLM. \verb+myQLM-fermion+ provides general tools for dealing with fermionic problems: representations of fermionic Hamiltonians (including usual Hamiltonians like the electronic structure Hamiltonian (Eq. \eqref{eq:Ham1}), the Hubbard and Anderson Hamiltonians), transformations from fermion operators to qubit operators (Jordan-Wigner, Bravyi-Kitaev, parity), generation of quantum evolution circuits via Trotterization, implementation of the quantum phase estimation algorithm, as well as various modules for variational algorithms such as VQE (see subsection \ref{subsec:VQE}): libraries of variational ansätze (like hardware-efficient ansätze, matchgate circuits, Low-Depth Circuit ansatz (LDCA) circuits), various optimization plugins (including the standard optimizers implemented in scipy\cite{2020SciPy-NMeth} (COBYLA, BFGS, etc) as well as the Simultaneous Perturbation Stochastic Algorithm (SPSA) and Particle-Swarm Optimizer (PSO)).


A submodule of \verb+myQLM-fermion+ \cite{myqlm-fermion} is specifically devoted to quantum chemistry. It provides tools for selecting active spaces (based on natural-orbital occupation numbers), generating cluster operators (and thus, via the aforementioned Trotterization tools, UCC-type ansätze), and initial guesses for their variational parameters. 
The architecture of QLM and \verb+myQLM-fermion+  allows for experts in a given field to construct their own advanced modules with the QLM building blocks. 

The key building block of quantum chemistry computations is the Hamiltonian, Eq. \ref{eq:Ham1}. On QLM, it is described by an object \verb+ElectronicStructureHamiltonian+

\begin{lstlisting}[language=Python]
from qat.fermion import ElectronicStructureHamiltonian
hamiltonian = ElectronicStructureHamiltonian(h, g)
\end{lstlisting}
where \verb+h+ and \verb+g+ are the tensors $h_{pq}$ and $h_{pqrs}$ of Eq. \eqref{eq:Ham1}.
Such an object also describes cluster operators such as the ones described in Eq.~\eqref{nexcitations} truncated at $x=2$. For instance, the following snippet
\begin{lstlisting}[language=Python]
from qat.fermion import get_cluster_ops
cluster_ops = get_cluster_ops(n_electrons, nqbits=nqbits)
\end{lstlisting}
creates the list containing the sets of single excitations $\lbrace c_k^\dagger c_i - c_i^\dagger c_k, k \in virt, i \in occ  \rbrace$ and double excitations $\lbrace c_l^\dagger c_k^\dagger c_j c_i - c_j^\dagger c_i^\dagger c_l c_k, k, l \in virt, i, j \in occ  \rbrace$.

These objects can be readily converted to a spin (or qubit) representation using various fermion-spin transforms:

\begin{lstlisting}[language=Python]
# Jordan-Wigner
from qat.fermion.transforms import transform_to_jw_basis
hamiltonian_jw = transform_to_jw_basis(hamiltonian)
cluster_ops_jw = [transform_to_jw_basis(t_o) for t_o in cluster_ops]

# Bravyi-Kitaev
from qat.fermion.transforms import transform_to_bk_basis
hamiltonian_bk = transform_to_bk_basis(hamiltonian)
cluster_ops_bk = [transform_to_bk_basis(t_o) for t_o in cluster_ops]
\end{lstlisting}
The transformed objects are now in the form of Eq. (\ref{Pauli-Hamiltonian}). With these qubit operators, one can then easily contruct a simple UCCSD ansatz via trotterization \text{red}{(as given Eq.~\eqref{tettor}) of the exponential of the parametric cluster operator defined as  \verb+cluster_ops_jw+}:
\begin{lstlisting}[language=Python]
from qat.lang.AQASM import Program, X
from qat.fermion.trotterisation import make_trotterisation_routine

prog = Program()
reg = prog.qalloc(nqbits)
# Create Hartree-Fock state (assuming JW representation)
for qb in range(n_electrons):
    prog.apply(X, reg[qb])

# Define the full cluster operator with its parameters
theta_list = [prog.new_var(float, "\\theta_{%s}" % i) for i in range(len(cluster_ops_jw))]
cluster_op = sum([theta * T for theta, T in zip(theta_list, cluster_ops_jw)]) 

# Trotterize the Hamiltonian (with 1 trotter step)
qrout = make_trotterisation_routine(cluster_op, n_trotter_steps=1, final_time=1)
prog.apply(qrout, reg)
circ = prog.to_circ()
\end{lstlisting}
The circuit we constructed, \verb+circ+, is a variational circuit that creates a variational wavefunction $|\psi(\theta)\rangle$. Its parameters can be optimized to minimize the variational energy $\displaystyle
E(\theta) = \sum_j h_j\langle\psi(\theta)|\prod_{i=0}^{n_o-1}\sigma_i^j|\psi(\theta)\rangle.$
This is done by a simple VQE loop:
\begin{lstlisting}[language=Python]
# create a quantum job containing the variational circuit and the Hamiltonian 
job = circ.to_job(observable=hamiltonian_jw, nbshots=0)

# import a plugin to perform the optimization
from qat.plugins import scipyMinimizePlugin
optimizer_scipy = scipyMinimizePlugin(method="COBYLA", tol=1e-3, options={"maxiter": 1000}, x0=theta_init)

# import a QPU to execute the quantum circuit
from qat.qpus import get_default_qpu

# define the quantum stack
stack = optimizer_scipy | get_default_qpu()

# submit the job and read the result
result = stack.submit(job)

print("Minimum energy =", result.value)
\end{lstlisting}

The simulated QPU used in the previous code snippet can be readily replaced by an actual QPU by using the \verb+qat-interop+ module. For instance, in the following snippet, we use a transmon QPU by IBM:
\begin{lstlisting}[language=Python]
from qat.interop.qiskit import BackendToQPU
qpu= get_default_qpu()
qpu = BackendToQPU(token=MY_IBM_TOKEN, ibmq_backend="ibmq_guadalupe")
\end{lstlisting}
Conversely, circuits created using other quantum programming frameworks can be converted to the QLM format. For instance, here we convert a circuit written in the Google Cirq format to a QLM circuit (that can then be fed to a QLM simulator):
\begin{lstlisting}[language=Python]
from qat.interop.cirq import cirq_to_qlm
qlm_circ = cirq_to_qlm(your_google_circ)
\end{lstlisting}


\section{The OpenVQE Package}\label{sec:OpenVQE}
OpenVQE facilitates the development of new algorithms on quantum computers. It consists of a couple of new modules that extend the main myQLM/myQLM-fermion implementations (see section \ref{sec:qlm} above). Those modules allow to build normal UCCSD algorithm, its variants and ADAPT-VQE algorithms as reviewed in section \ref{sec:review}. To explain in another fashion, OpenVQE allows to perform calculations using new types of UCC methods that are different from the QLM predefined ones such as the regular UCCSD ansatz. 

OpenVQE consists of two main modules that are inside the "openvqe" folder:

\begin{figure*}[h!]
\centering
\includegraphics[width=0.85\textwidth]{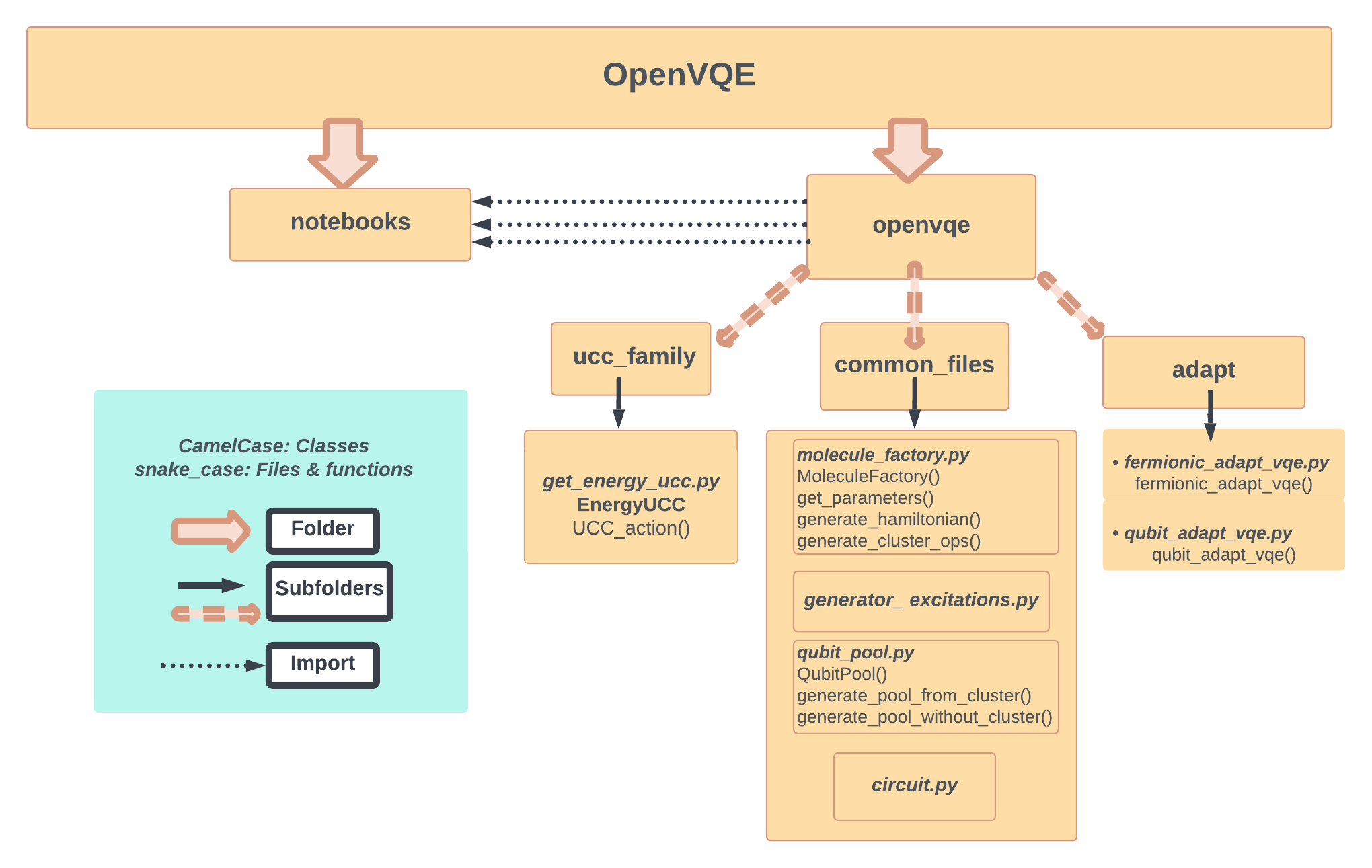}
\caption{Flow chart of the OpenVQE package. The code is given in our Github repository and documentation \cite{openvqe}.}
\label{fig:flow_chart_open_vqe}
\end{figure*}

\begin{itemize}
    \item \textbf{UCC Family} denoted in the code by "ucc\_family": this module includes different classes and functions to generate the fermionic cluster operators (fermionic pool) and the qubit pools, and to get the VQE optimized energies in the cases of active and non-active orbital selections.
    
    \item \textbf{ADAPT} denoted by "adapt": it includes two sub-modules: 
    \begin{itemize}
        \item \textbf{Fermionic-ADAPT}: containing functions performing the fermionic-ADAPT-VQE algorithmic steps in the active and non-active space selections;
        \item \textbf{Qubit-ADAPT}: containing functions that perform the qubit-ADAPT-VQE algorithmic steps calculation in the active and non-active space orbital selections.
    \end{itemize}
\end{itemize}
Additionally, "openvqe" contains a subfolder named "common\_files" that stores all the internal functions needed to be imported for executing the two modules. 
OpenVQE also contains a "notebooks" folder that allows the user to run and test the above two modules (which are theoretically described in the review section \ref{sec:review}).
A brief sketch of the OpenVQE implementation  is explained in the flow chart presented in Figure \ref{fig:flow_chart_open_vqe}. 
We remind  the reader that we will  describe below the code of OpenVQE for the full space selections and some internal function examples  such as those related to "generator\_excitations" and "fermionic\_adapt\_vqe. The codes related to active space selections can be found in the GitHub website \cite{openvqe} inside "notebooks" with prefix "active\_space".

In order to develop a new algorithm using our OpenVQE package, it is important to list the different parameters and return variables contained in its main functions.  

The first crucial parameters are the ones specifying the characteristics of the molecules such as the symbol of the molecule;  the type of excitation generator (generators to produce the excitations, see sec. \ref{UCCsection}):  UCCSD (Eq. \ref{uccsdtrotter}), QUCCSD (Eq. \ref{quccsdtrotter}), UCCGSD (Eq. \ref{uccgsd}), K-UpCCGSD (Eq. \ref{kupccgsd}) and spin-complemented pair (Eqs. S1 and S2 in the Supplementary Material\cite{supplementary});  the spin transformation mapping (JW, Bravi-Kitaev, parity basis), choice of active or non-active space selection, denoted respectively in our code as \verb+molecule_symbol+, \verb+type_of_generator+, \verb+transform+, and \verb+active+.
\begin{lstlisting}[language=Python]
# molecule examples: H2, H4, H6, LiH, H2O, CO, CO2, NH3 etc... 
molecule_symbol = 'H2O'
# suppose user choose the spin-complemented generalized singlet and doublet  denoted as spin_complement_gsd
type_of_generator = 'spin_complement_gsd'
# user can type other type of generators, such as: uccsd, quccsd, uccgsd,  k-upccgsd, etc... 
# we type Jordan wigner (JW),  user can  also type Bravyi-Kitaev or Parity basis
transform = 'JW'
# no active space selection, set: 
active = False
# for active space (AS), switch to  "active = True"
\end{lstlisting}

 The user specifies these parameters in a class called \verb+MoleculeFactory+. This class takes those parameters as input: 
\begin{lstlisting}[language=Python]
from openvqe.common_files.molecule_factory import MoleculeFactory

# returns the properties of a molecule: 
r, geometry, charge, spin, basis = MoleculeFactory.get_parameters(molecule_symbol = 'H2O')
\end{lstlisting}
define a function named as \verb+generate_hamiltonian+ that generates the electronic structure Hamiltonian (\verb+hamiltonian+) and other properties such as the spin hamiltonian (for example \verb+hamiltonian_jw+), number of electrons (\verb+n_els+), the list contains the number of natural  orbital occupation numbers (\verb+noons_full+), the list of orbital energies (\verb+orb_energies_full+) where the orbital energies are doubled due to spin degeneracy and \verb+info+ which is a dictionary that stores energies of some classical methods( such as Hartree-Fock, CCSD and FCI):
\begin{lstlisting}[language=Python]
Hamiltonian, hamiltonian_jw, n_els, noons_full, orb_energies_full, info = MoleculeFactory.generate_hamiltonian(molecule_symbol='H2O', active=False, transform='JW')
\end{lstlisting}
In addition to that, we define another function named as \verb+generate_cluster_ops()+ that takes as input the name of excitation generator user need (e.g., UCCSD, QUCCSD, UCCGSD, etc.) and internally it calls the file name \textit{generator\_excitations.py} which allows \verb+generate_cluster_ops()+ to return as output the size of pool excitations, fermionic operators, and  JW transformed operators denoted in our code respectively as pool\_size, \verb+cluster_ops+ and \verb+cluster_ops_jw+:
\begin{lstlisting}[language=Python]
from .generator_excitations import (uccsd, quccsd, singlet_gsd, singlet_sd,singlet_upccgsd, spin_complement_gsd, spin_complement_gsd_twin)
pool_size,cluster_ops,cluster_ops_jw = MoleculeFactory.generate_cluster_ops(molecule_symbol='H2O',  type_of_generator='spin_complement_gsd', transform='JW', active=False)
# in our example 'spin_complement_gsd':
def generate_cluster_ops():
    pool_size,cluster_ops,cluster_ops_jw= None, None, None
    if type_of_generator == 'spin_complement_gsd':
        pool_size,cluster_ops,cluster_ops_jw   = spin_complement_gsd(n_el,n_orb,'JW')
#  elif for other excitations (uccsd, quccsd, singlet_upccgsd...) 
#    ::::
    return  pool_size,cluster_ops,cluster_ops_jw       
\end{lstlisting}
Once these are generated, we import them as input to the UCC-family and ADAPT modules.\\
In the example of fermionic-ADAPT sub-module, we call the function \verb+fermionic_adapt_vqe()+ that takes as parameters the fermionic cluster operators, spin Hamiltonian, maximum number of gradients to be taken per iteration, the type of optimizer, tolerance, threshold of norm ($\epsilon$) and the maximum number of adaptive iterations:
\begin{lstlisting}[language=Python]
from openvqe.adapt.fermionic_adapt_vqe import fermionic_adapt_vqe

# choose maximum number of gradients needed (1,2,3....)
n_max_grads = 1
# choose optimizer needed (COBYLA, BFGS, SLSQP, Nelder-Mead etc...)
optimizer = 'COBYLA' 
tolerance = 10**(-6)  
# according to a given norm value we stop the ADAPT-VQE loop
type_conver = 'norm'
threshold_needed = 1e-2
# the maximum external number of iterations to complete the ADAPT-VQE under a given threshold_needed
max_iterations = 35
fci = info['FCI']
# sparse the Hamiltonian and cluster operators using myQLM-fermion tools obtained from MoleculeFactory, which are explicitly:
hamiltonian_sparse = hamiltonian_jw.get_matrix(sparse=True)
cluster_ops_sparse = cluster_ops.get_matrix(sparse=True)
# reference_ket  and hf_init_sp can be obtained from class MoleculeFactory():
reference_ket, hf_init_sp = MoleculeFactory.get_reference_ket(hf_init, nbqbits, 'JW')
# when all these parameters are satisfied, then fermionic-ADAPT-VQE function is:
fermionic_adapt_vqe(cluster_ops, hamiltonian_sparse, cluster_ops_sparse, reference_ket, h_sp, cluster_ops_jw, hf_init_sp, n_max_grads, fci, optimizer,  tolerance,type_conver = type_conver, threshold_needed = threshold_needed, max_external_iterations = max_iterations)
\end{lstlisting}
\verb+fermionic_adapt_vqe()+ calls internally other functions allowing the execution of the steps from 1th to 4th given in subsection \ref{ADaPT-VQE_review}: (1) prepare the trial state through \verb+prepare_state()+; (2) compute analytically  the commutator between the hamiltonian and the fermionic operator through  \verb+compute_gradient()+ (or numerically by using the command \verb+hamiltonian_jw | cluster_ops_sp+ ); (3) collect and arrange the gradients in a list from maximum to minimum (with avoiding the zeros) through \verb+sorted_gradient()+; (4) if the norm is less than the ($\epsilon$), the program exits returning the generated values, else it will continue calculating the maximum gradient(s) of operator(s) depending on the number of maximum gradient in the input, after that, we append the operator(s) associated with their parameter(s) to the left of the previous trial state (step (1)) during which we apply VQE using function \verb+ucc_action()+,  in order to optimize the parameter(s) until satisfying the threshold's condition;  
\verb+fermionic_adapt_vqe()+ returns at the end: (i) the number of classical parameters for the final ansatz, (ii) the number of CNOT gates in the ansatz circuit, (iii) number of other gates, (iv) list of optimized energies corresponding to external iterations and (v) energy subtracted from the full configuration interaction (FCI).
 The qubit-ADAPT sub-module is globally similar to the fermionic-ADAPT in terms of the code structure, some key steps are different and makes it unique in its nature. Those steps can be summarized in the following sequence: (i) we use qubit pool generators using \verb+QubitPool+  class instead of fermionic ones; (ii) the preparation of the trial state is different from the fermionic-ADAPT one; (iii) the gradients calculation is not the same. It returns the same properties as that of fermionic-ADAPT-VQE.


\section{Systematic Benchmarking of the Performances of the UCC and ADAPT-VQE Algorithms on Several Molecules using the OpenVQE Package}\label{sec:adaptvqe}
In this section, we used OpenVQE to perform numerical estimations of ground state energies of various molecules using the VQE-UCC-family and ADAPT-VQE modules. The studied molecules require a number of qubits ranging from 4 to 24 (see Figure \ref{fig:TimingQLM}).
We perform only noiseless simulations in order to minimize computational runtime.
All simulations were performed using TotalEnergies's in-house HPC computing platform \footnote{The server we used for our calculations has the following properties: it is Linux machine named "skelling", with 162 cores, each core containing 2 threads and a total memory of  \SI{3094000}{\mebi\byte}.}. We validated our results by comparing some of them with other works obtained with similar methodologies.
For each numerical simulation in the UCC-family or ADAPT-VQE modules, we use the following common functions implemented in OpenVQE to calculate:  (i) the molecular orbital integrals using the PYSCF package\cite{sun2018pyscf} with the STO-3G, 6-31G and cc-pVDZ basis sets;
(ii) the molecular Hamiltonian mapping, obtained by applying the Jordan-Wigner transformation;
(iii) the ansatz circuits, based on the CNOT staircase method, except for QUCCSD version for which we use the two circuits representing the single and double qubit evolutions (see Figure 1 and 2 in \cite{Yordan2021}, respectively). For optimization,
we used both optimzers BFGS and SLSQP from the \textit{scipy.optimize} library\cite{scipy}.We also set the commonly accepted threshold for chemical accuracy, which is 1 kcal/mol  (i.e 1.59$\times 10^{-3}$ Hartees(Ha)). This accuracy represents the error
difference between the predicted energy and FCI energy. We use this value as a standard reference in all our results.\\

In order to show how openVQE performs, we split our results into four main subsections:
(i) subsection \ref{simulator} displays the timings of the QLM simulator for constructing the UCCSD ansatz on a simulator as well as for obtaining the expectation value of the electronic structure Hamiltonian for a set of molecules (STO-3G basis set) ranging from 4 to 24 qubits;
(ii) subsection \ref{UCC-results}, dedicated to the UCC-family module, describes the performance of UCCSD-VQE for H$_6$ with the STO-3G basis set and for LiH using 6-31G basis set, with different sizes of active spaces, using Møller-Plesset wavefunction in second order (MP2 pre-screening) as initial guess;
(iii) subsection \ref{Open-VQE-ADAPTresults}  uses the ADAPT-VQE module. It describes the comparison between the fermionic and qubit-ADAPT-VQEs, in terms of the number of operators as well as the chemical accuracy, by benchmarking the three molecules H$_4$, LiH and H$_6$, in the STO-3G orbital basis set,  by assuming a full space selection approximation in each molecule.
(iv) subsection \ref{comparison-tables} introduces a brief Table summarizing the comparison of UCC-family methods involving fermionic excitations (UCCSD and UCCGSD) with the fermionic-ADAPT-VQE approach. Several molecules are considered and the results are discussed in terms of number of parameters, CNOT counts, chemical accuracy, and computational time.\\

In addition to the results presented below in this section, there are also additional interesting results obtained using the OpenVQE UCC family module. Indeed,  we simulated several other test molecules, using a larger number of qubits and different chemical basis sets assuming both full and active space selections.
Such computations were performed in order to demonstrate further the capabilities of our module to perform large calculations and to reach chemical accuracy despite the molecular size increase.
These results are detailed in Section 2 of the Supplementary Materials\cite{supplementary}.
\subsection{The Standard UCCSD Ansatz: Representative Computational Performances with OpenVQE}\label{simulator}
To illustrate the performance of our OpenVQE package on CPUs, we chose to use a standard laptop\footnote{laptop properties are: Intel(R) i5-10310U, Clock speed 2.21 GHz processor, OP windows 10.} to evaluate the time to solution required to perform two operations that are common to most of the variational algorithms:
(i) the application of the ansatz;
(ii) the computation of the expectation value of a molecular Hamiltonian or another observable.
Examples of such operations are: the implementation of a hardware efficient ansatz, or a unitary coupled cluster ansatz. Observables are for example the electronic structure Hamiltonian, a commutator operator, Pauli string terms etc, which  can be applied with either of these ansätze to  measure their expectation value.
As a test, we chose here the UCCSD method to evaluate the timing associated to various computations. It is important to note that, in practice, for the particular case of the simple UCCSD, OpenVQE and the native QLM (QLM simulator version: 1.2.1) timings are identical.
To measure such timings, we conducted benchmarks with a set of molecules using OpenVQE and the STO-3G basis set (H$_2$, H$_4$, LiH, H$_2$O, NH$_3$, CH$_4$, CO, HCN, and C$_2$H$_2$). 

\begin{figure*}[h!]
     \centering
     \subfloat[]{\includegraphics[width=0.41\textwidth]{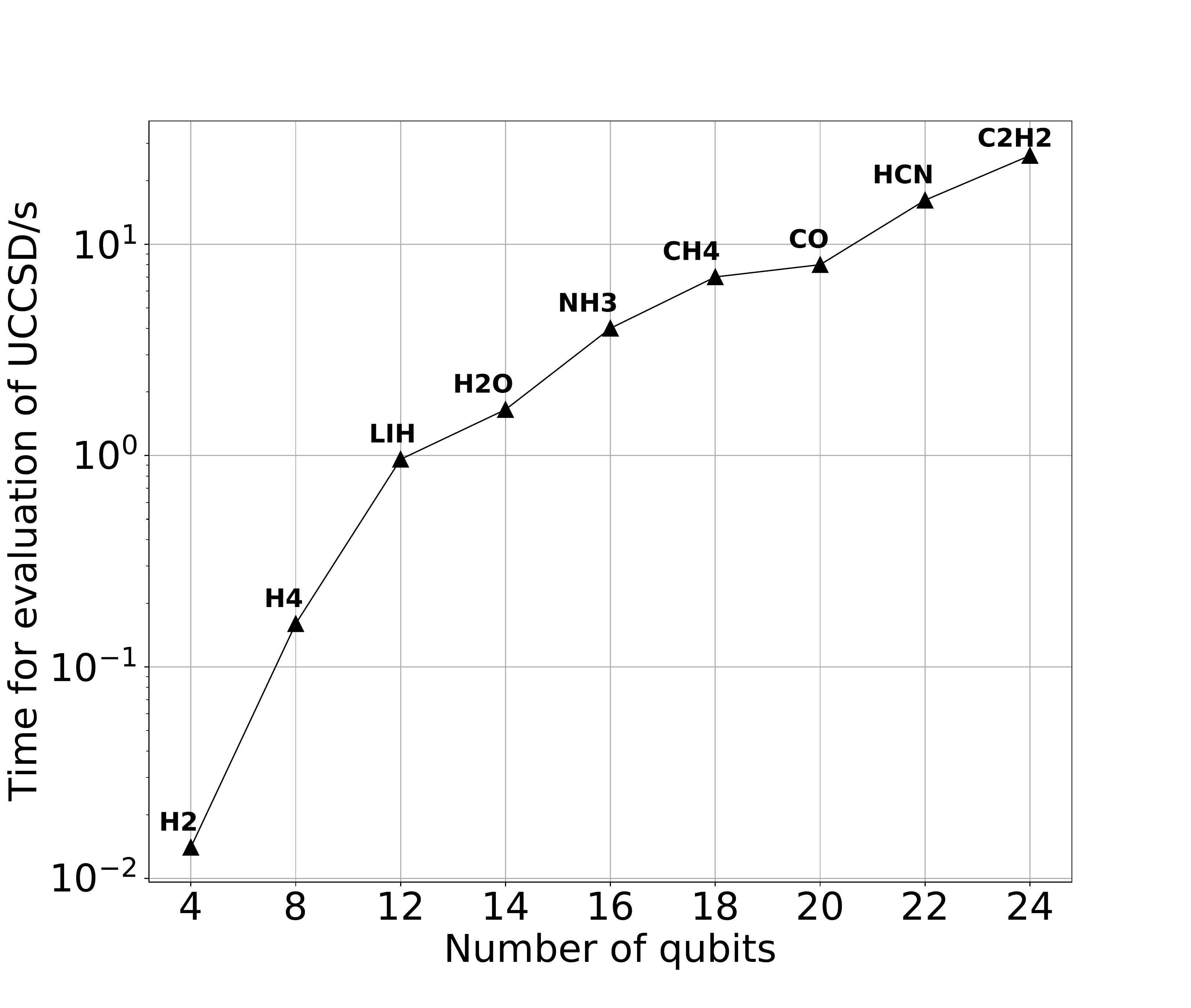}\label{<figure1>}}
     \subfloat[]{\includegraphics[width=0.41\textwidth]{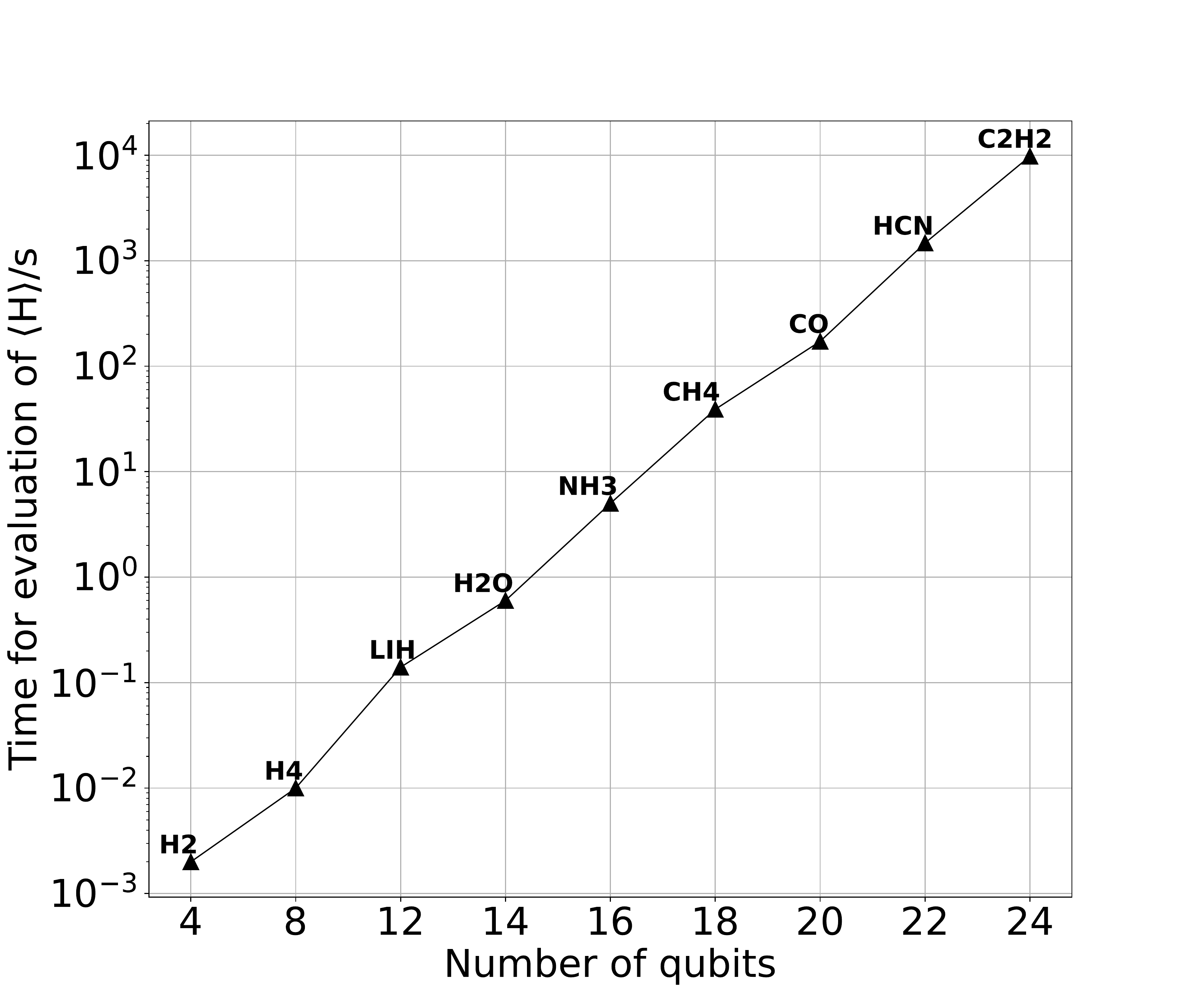}\label{<figure2>}}
     \par\medskip
     \subfloat[]{\hspace*{-1cm}\includegraphics[width=0.62\textwidth]{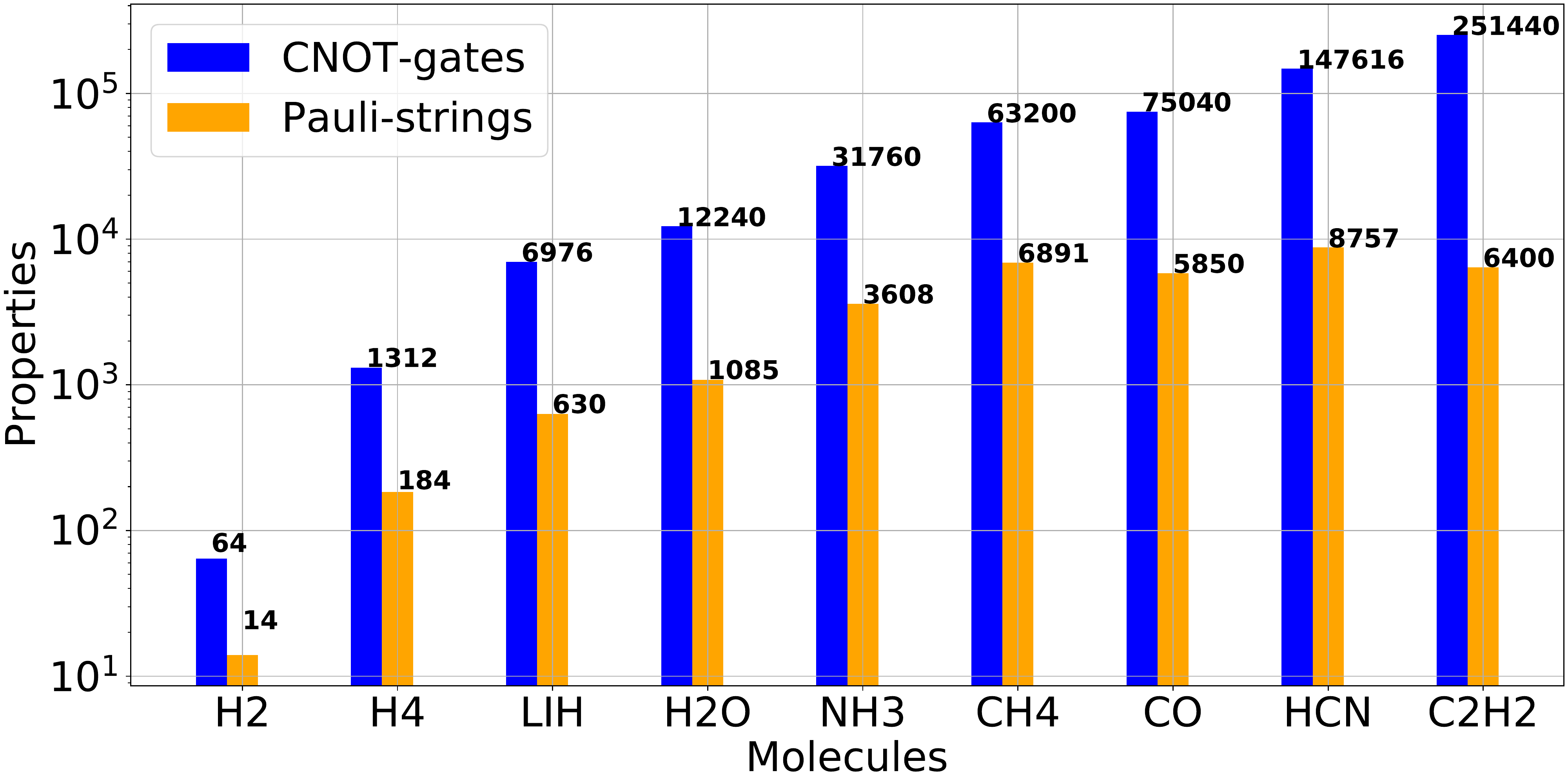}\label{<figure3>}}
     \caption{Panel(a): timings for the application of a UCCSD
circuit by OpenVQE while increasing the number of qubits. Panel (b): timings for the evaluation of increasing size molecular Hamiltonians (summed over Pauli strings and using using JW transformation) for the following molecules: H$_2$, H$_4$, LiH, H$_2$O, NH$_3$, CH$_4$, CO, HCN, C$_2$H$_2$ . Panel (c) shows the number of CNOT gates required to complete the UCCSD circuit
and the number of Pauli strings in the Hamiltonian. }
     \label{fig:TimingQLM}
\end{figure*}

Figure \ref{fig:TimingQLM} (a) displays the timings associated to the generation of a  state-vector corresponding to the UCCSD ansatz (see Eq. \ref{uccsdtrotter}) from OpenVQE. The computational cost for this generation grows sub-exponentially with the number of qubits.
We observe that the package can compute the UCCSD wavefunction within less than a second for a molecule requiring  4 to 12 qubits, between a second and 10 seconds for molecules associated to a 14-20 qubits range and up to a few minutes for larger molecules (22-24 qubits).
Figure \ref{fig:TimingQLM} (b) represents the timings required to measure the expectation value of the JW transformed Hamiltonian (see Equation \ref{Pauli-Hamiltonian})  of each of these molecules using UCCSD ansatz. This measurement increases linearly with the number of qubits and is performed using the UCCSD circuit together with the "Job" class which will be submitted to QPU to obtain the expectation value.  The process is described by the code function called \verb+UCC_action()+ which can be found in the UCC module openVQE\cite{openvqe}.
We observe that the computational cost of performing the measurement increases exponentially with the number of qubits. 
In particular, we notice that molecular systems up to 14 qubits like H$_2$O (10 electrons) using OpenVQE, can be performed within a second for a UCCSD circuit. This circuit involves 12,240 CNOT gates and 1,083 Pauli strings, as shown in Figure \ref{fig:TimingQLM}(c). This Figure also shows the details of the number of CNOT gates and Pauli strings associated to each molecule.

\subsection{UCCSD: Active Space Selections and MP2 Pre-screening}
\label{UCC-results}
\begin{figure*}[h!]
     \centering
     \subfloat[][Errors for the H$6$ molecule ground state ]{\includegraphics[width=0.5\textwidth]{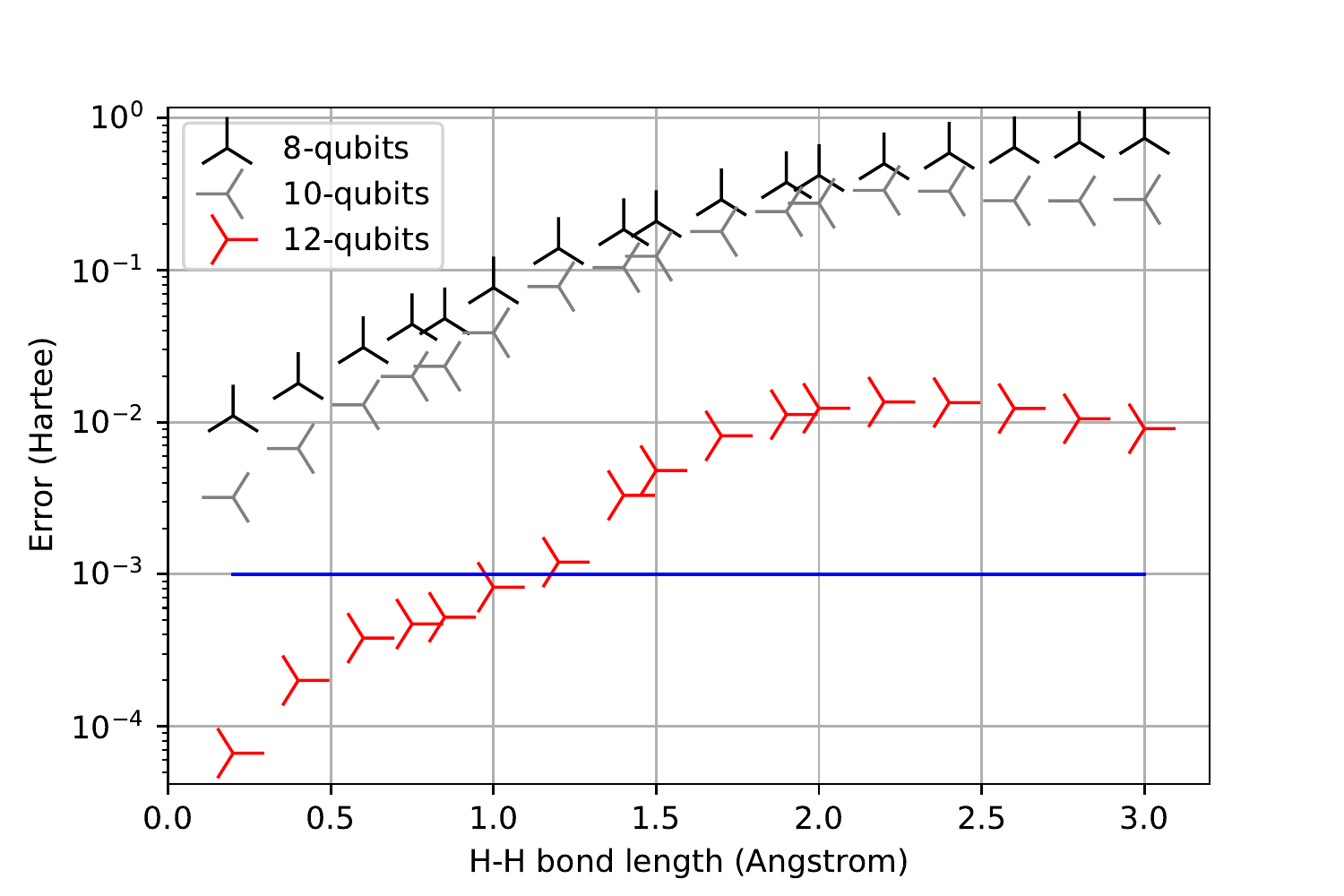}}
     \subfloat[][Errors for the LiH molecule ground state.]{\includegraphics[width=0.5\textwidth]{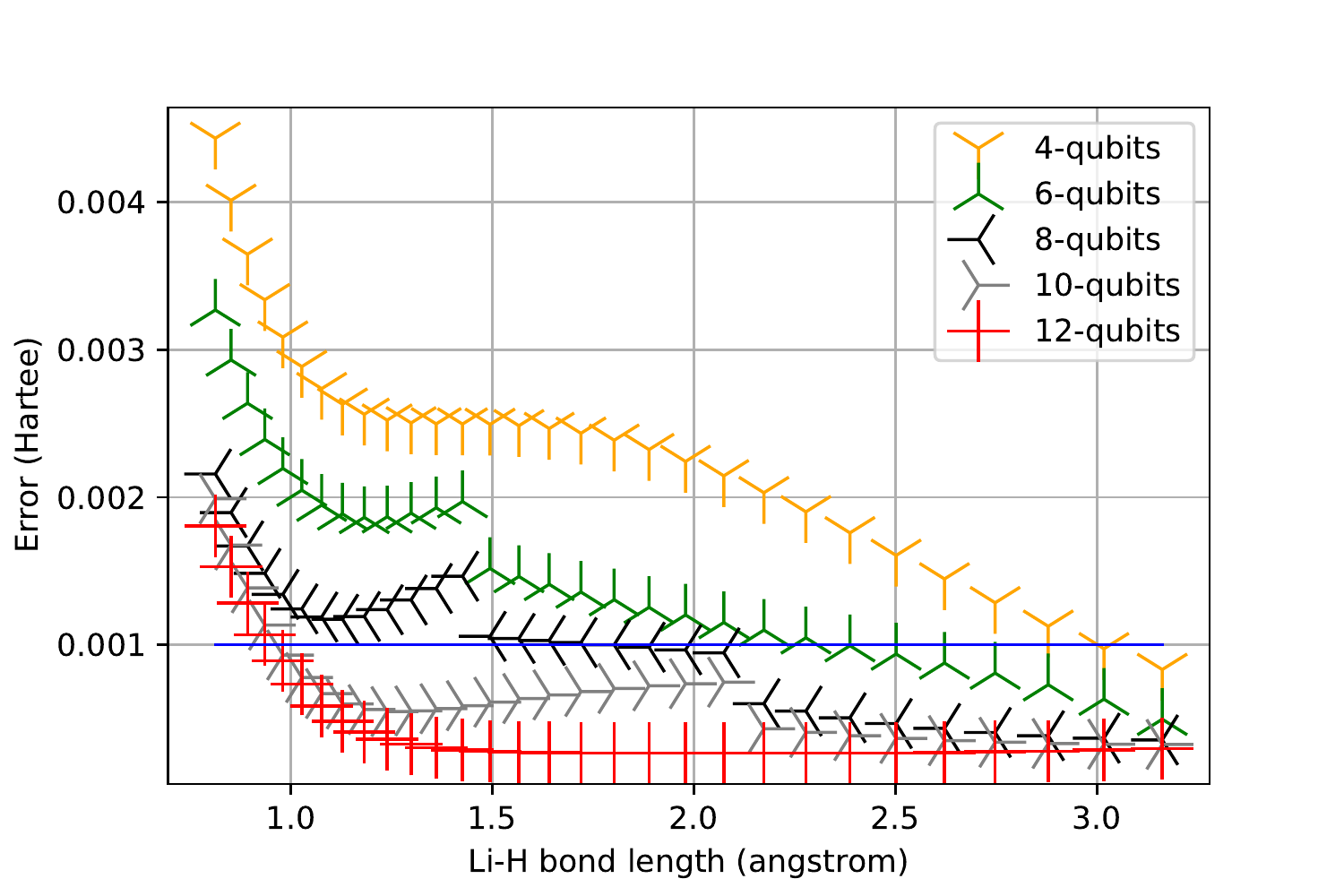}}
     \caption{VQE results for the H$_6$ molecule using the STO-3G basis set (a) and for LiH with the 6-31G basis (b). The dissociation profile is showed, using different active space selections and are calculated by using the first order trotterization UCCSD ansatz. The size of qubits represents the number of active spin-orbitals. The blue line indicates the chemical accuracy.}
     \label{fig:groundStates}
\end{figure*}

We test the active space (AS) selection approach by simulating  H$_6$ linear geometry type within STO-3G basis set and LiH within 6-31g basis set, for a range of bond lengths for both molecules.  We use MP2 pre-screening as initial guesses in the BFGS optimizer.
For the H$_6$ molecule, which has a Hilbert space spanned by 12 Hartree-Fock orbitals, 6 occupied and
6 virtual (i.e., unoccupied) we choose two different active spaces:
(i) First 2 spin orbitals where the lowest energies are always filled and  2 spin orbitals where the  highest energies are always empty, as these 4 spin-orbitals are not considered in the active space selection, the system consists of 8 active spin orbitals with 4 active electrons, which means a 8-qubit Hamiltonian;
(ii) when we consider that only the first two spin-orbitals are always filled, the system  has then 4 active electrons and 10 spin-orbitals corresponding to 10 qubit Hamiltonian. Figure \ref{fig:groundStates}(a) shows the error of ground state energies of the H$_6$ molecule at different dissociation profiles, with different sizes of the active spaces ranging from a minimum of 8 qubits to the full space of 12 qubits.
This error corresponds to absolute value of VQE-UCCSD results substracted from FCI energy. By increasing the active space from 8 to 10 qubits, we observe only little improvement of the accuracy and both sizes are still far from approaching the chemical accuracy (see the blue line which was set at 10$^{-3}$ Ha).
However in the full space case, the VQE results approach the chemical accuracy before a bond length equal to 1.5 \AA.
Interestingly, for bond length $\geq 1.5$ \AA, even with full space, the error deviates well beyond the chemical accuracy but it seems to be curving again toward the blue line beyond 3.0 \AA. This error means that triple excitations are probably required. 
Results comparable to our findings for H$_6$ can be seen in Figure 7 of \cite{Xia}, but this work only considered the non-active case. 
Figure \ref{fig:groundStates}(b) shows the same energy profile but for the LiH molecule.
Using the 6-31G basis set, LiH has a Hilbert space spanned by 22 HF orbitals, 4 occupied and 18 virtual (i.e., unoccupied).
But five active space selections are chosen, leading to 4, 6, 8, 10 and 12 qubit Hamiltonians. For all choices of active spaces here, the error decreases steadily as the the bond length increases.
The 10 and 12 qubit Hamiltonians are below the chemical accuracy for bond length $>1.0$\AA, while the largest deviation measured is in the intermediate bond lengths range between 1 and 2.4 \AA.
Interestingly, the 8, 10 and 12 qubit results fall within a narrow range, at bond length before 1.0 and at bond length beyond 2.2\AA. Surprisingly, we noticed some jumps in the accuracy for the three cases 6, 8 and 10 qubits at some bond lengths, for example around 2.2\AA. 
A similar analysis has been made in \cite{barkoutsos2018quantum} for two different molecules than ours, by changing the number of qubits in different basis set: (6-31G) in H$_2$ and (STO-3G) in H$_2$O. In particular they observed similar jumps we also observed for H$_2$O at intermediate range, which as mentioned there it might correspond to geometries close to the so-called Coulson-Fisher point where spin-symmetry breaking can occur\cite{gunnarsson1976exchange}. 


\subsection{Performance of ADAPT-VQE Module in OpenVQE}
\label{Open-VQE-ADAPTresults}
\begin{figure*}[ht!]
    \centering
    \includegraphics[width=1\textwidth]{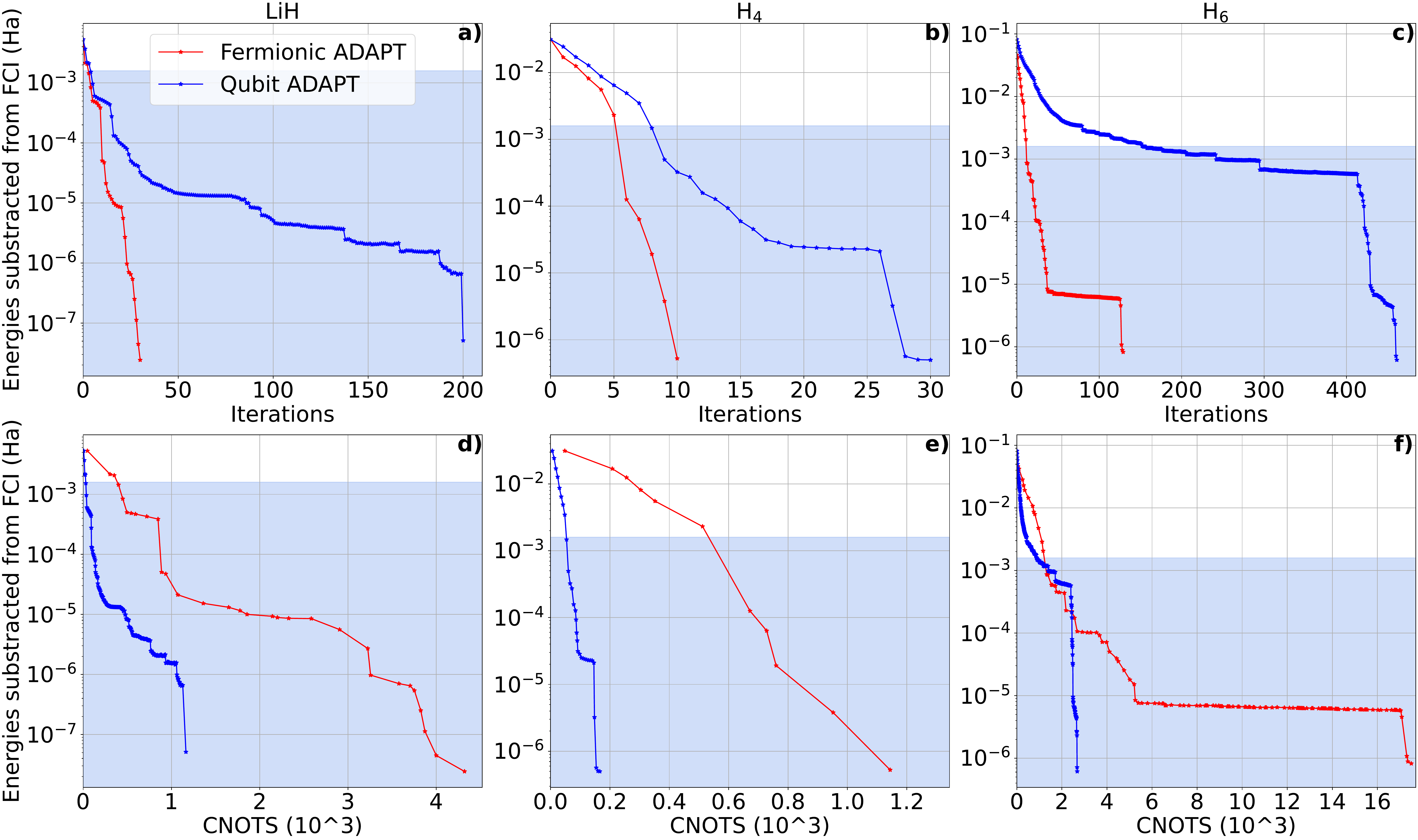}
    \caption{Energy convergence plots for the near equilibrium ground states of LiH, H$_4$ and H$_6$ using the STO-3G orbital basis set, at bond distances $r_{\text{Li-H}}$ = 1.45\AA, $r_{\text{H-H}}$ = 0.85\AA, and $r_{\text{H-H}}$  = 1.0\AA. The red plots are obtained with a fermionic-ADAPT-VQE algorithm and the blue plots are obtained with qubit-ADAPT-VQE. Top panels (a,b,c) present the energy accuracy as a function of the ansatz parameters (note that for both ADAPT-VQE, the number of parameters are the same as the number of operators taken per iteration). Bottom panels (d,e,f) present the energy accuracy as a function of the CNOT count of the ansatz circuit. The blue area corresponds to the chemical accuracy.}
    \label{fig:H6}
\end{figure*}

In Figure \ref{fig:H6}, we represent the energy convergence for LiH, H$_4$ and H$_6$, which are (8-12) qubit systems, obtained with the fermionic-ADAPT-VQE and qubit-ADAPT-VQE algorithms. The three molecules are at bond lengths $r_{\text{Li-H}}$= 1.45\AA,  $r_{\text{H-H}}$ = 0.85\AA, and H$_6$ at
bond distance $r_{\text{H-H}}$ = 1.0 \AA, respectively.  
All convergence plots are terminated when the norm value of the gradient is below the threshold $\epsilon$ (see  section \ref{ADaPT-VQE_review}), which is between $10^{-3}$ and $10^{-4}$
for fermionic-ADAPT-VQE,  and  between $10^{-3}$ and $10^{-6}$ for qubit-ADAPT-VQE, depending on the molecule.
Since we find SLSQP and BFGS optimizers behave comparably for energy convergence (after using MP2 initial guess) as is noticed in Figure. S3 in the Supplementary Material\cite{supplementary}, we used  SLSQP optimizer for fermionic-ADAPT-VQE and the BFGS optimizer for qubit-ADAPT-VQE.
The plots in Figure \ref{fig:H6} present the two ADAPT algorithms in terms of two important conditions, required to construct an ansatz and to achieve a specific chemical accuracy :
(i) the number of iterations (or parameters);
(ii) the number of CNOT gates.
Figures \ref{fig:H6} (a,b,c)- show that the fermionic-ADAPT-VQE converges faster than the qubit-ADAPT-VQE, requiring systematically fewer number of iterations and thus variational parameters. However as it is shown in Figures \ref{fig:H6} (d,e,f)- the  CNOT counts from the fermionic-ADAPT-VQE required to reach convergence, are higher by more than one-order of magnitude compared to qubit-ADAPT-VQE.
Such results make sense with the fermionic-ADAPT-VQE approach since the type of operators is spin-complemented pairs (see Eqs. S1 and S2 in the Supplementary Material). Indeed, the JW transformation of single and double operators will lead to sets of two and eight Pauli strings, respectively, where each Pauli string is a tensor product of Pauli matrices.
According to theory, there are four CNOT staircases (for a single excitation) and sixteen (for a double excitation)\cite{Yordan2020}.
The qubit-ADAPT-VQE anszatz (see section \ref{ADaPT-VQE_review}) states that the pool consists of  many individual Pauli strings. This means the qubit-ADAPT-VQE needs more ansatz elements (i.e operators or parameters) to reach convergence.  However, at the same time, the CNOT counts are strongly reduced compared to fermionic-ADAPT-VQE.
Similar CNOT counts results, number of parameters and chemical accuracy for fermionic-ADAPT-VQE involving LiH and H$_6$ molecules are presented in Figure 2 of \cite{grimsley2019adaptive}.
For the LiH molecule, we match these results. For H$_6$ some differences occur since we have converged our computations beyond the 10$^{-3}$ threshold. This added additional parameters to the ansatz, providing therefore a better accuracy.
In \cite{tang2021qubit} (Figure 1) results of fermionic and qubit-ADAPT-VQE involving LiH ($r= 2.0$\AA), H$_4$ ($r= 1.5$\AA) and 
$H_6$ ($r=2.0$\AA) are  also comparable to our findings despite some small differences are noted due to slightly different geometries in the three molecules, and convergence threshold.

\subsection{Comparison of Fixed-length Ansätze with ADAPT-VQE Methods}\label{comparison-tables}
In this subsection, we want to determine how many parameters (i.e. CNOT gates) are needed and which energy accuracy is reached by the "fixed-length ansätze" as opposed to the fermionic-ADAPT-VQE. To do so, we choose the LiH (1.45\AA) and H$_6$(1.0\AA) molecules as benchmarks.
In Figures \ref{fig:H6}(a,b,d and f), and in Figures (S4.a and S4.c in the Supplementary Material \cite{supplementary}), the attempt to reach chemical accuracy is displayed alongside with the number of CNOTs for both the fermionic-ADAPT-VQE and UCCSD, respectively.
The UCCGSD (see Eq. \ref{uccgsd}) method remains to be characterized. We do so using the BFGS optimizer in a similar VQE loop.
Table \ref{TableLiH} and Table \ref{TableH6} summarize the results obtained from calculations with each of the methods for LiH and H$_6$, respectively.
 \begin{table}[h]
\begin{tabular}{cccccc}
\rowcolor[HTML]{C0C0C0} 
\textbf{Method} & \textbf{Parameters} & \textbf{CNOT-gates} & \textbf{Energy (Ha)} & \textbf{Error (Ha)} & \textbf{Computational time} \\
UCCSD & $44$ & $6080$ & $-7.880973466$ & $8.85\times 10^{-6}$ & 1 hr, 11 min \\
\rowcolor[HTML]{EFEFEF} 
UCCGSD & $330$ & $97280$ & $-7.880973130$ & $9.18\times 10^{-6}$ & 1 day, 12 hr \\
Fermionic-ADAPT-VQE & $30$ & $4624$ & $-7.880982281$ & $2.43\times 10^{-8}$ & 2 hr, 43 min 
\end{tabular}
\caption{LiH ($r_{\text{Li-H}}$) = 1.45\AA  \ molecule at the STO-3G basis set level. Comparison between UCCSD and UCCGSD fixed-length ansätze with fermionic-ADAPT-VQE in terms of number of parameters, CNOT-gates and energies. Accuracy is displayed as the absolute value of the difference between the ansatz Energy and the FCI energy in Hartree (Ha).}
\label{TableLiH}
\end{table}
\begin{table}[h]
\begin{tabular}{cccccc}
\rowcolor[HTML]{C0C0C0} 
\textbf{Method} & \textbf{Parameters} & \textbf{CNOT-gates} & \textbf{Energy (Ha)} & \textbf{Error (Ha)} & \textbf{Computational time} \\
UCCSD & $54$ & $8544$ & $-3.235451570$ & $6.14\times 10^{-4}$ & 3 hr, 10 min \\
\rowcolor[HTML]{EFEFEF} 
UCCGSD & $330$ & $97280$ & $-3.2360518902$ & $1.43 \times 10^{-5}$ & 5 days, 7hr\\
Fermionic-ADAPT-VQE & $130$ & $21648$ & $-3.236065546$ & $7.33\times 10^{-7}$ & 7 day, 7 hr 
\end{tabular}
\caption{Linear H$_6$ ($r_{\text{H-H}}$) = 1.0\AA\ molecule at the STO-3G basis set level. Comparison between UCCSD and UCCGSD  fixed-length ansätze and fermionic-ADAPT-VQE in terms of number of parameters, CNOT-gates and Energy. Accuracy is displayed as the absolute value of the difference between the ansatz Energy and the FCI energy in Hartree (Ha).}
\label{TableH6}
\end{table}
We observe from Table \ref{TableLiH} that fermionic-ADAPT-VQE brings the most accurate values for the calculated energy by around two orders of magnitude compared with that of UCCSD and UCCGSD. A comparable accuracy between UCCSD and ADAPT-VQE occurs when a norm threshold $10^{-2}$ is used in ADAPT-VQE (which could be also noticed in figure 2.b of \cite{grimsley2019adaptive}).  This means that as the norm threshold decreases, the ADAPT-VQE approach reaches a better accuracy.   
Remarkably, the number of parameters/CNOTs (30/4624) is still small compared to UCCSD and UCCGSD even though the considered threshold is 10$^{-3}$. 
Results displayed in Table \ref{TableH6} demonstrate again that ADAPT-VQE yields to the most accurate energy for the H$_6$ molecule, by at least two orders of magnitude as compared to UCCGSD and UCCSD. Although UCCSD requires a fewer number of parameters and CNOTs than with ADAPT-VQE, the latter brings better accuracy. This is again due to the fact that we controlled the norm threshold to  10$^{-4}$. A comparable accuracy between UCCSD and ADAPT-VQE in H$_6$ occurs when a norm threshold $10^{-1}$ is used in ADAPT-VQE as noticed in Figure 2(h) of reference \cite{grimsley2019adaptive}). It shows the fact that, at this level of threshold, ADAPT-VQE decreases the counts of CNOTs and parameters.
For UCCGSD, its energy lies between UCCSD and ADAPT-VQE ones, while its number of CNOTs and parameters remain very high.\\ With those results, one can deduce that ADAPT-VQE brings more accurate results than UCCSD by several order of magnitudes, depending on how the user can control the threshold value, but the cost associated to the use of ADAPT-VQE with and increased threshold is linked to the associated higher number of iterations and CNOTs. These latter increase with the number of qubits, which consequently leads to higher computational requirements in term of gradient computations which could be a practical problem on a real QPU.
For UCCGSD (together with UCCSD), the analysis of numerical simulations applied to  H$_4$, H$_2O$ and N$_2$ were shown in reference \cite{lee2018generalized}, illustrating the fact that although the UCCGSD ansatz yields very accurate results, the problem in using this method is linked to its high circuit depth. This is the opposite to the UCCSD case (or even k-UpCCGSD with at least k>2, see \cite{lee2018generalized}).
\\
\section{Conclusion and Future Work}
In this article, we discussed in detail our newly introduced open source package, OpenVQE. OpenVQE allows one to perform quantum chemistry computations based on the Quantum Learning Machine (QLM) thanks to the newly introduced myQLM-fermion open-source library that gather the key QLM resources that are important for quantum chemical developments. The OpenVQE/myQLM-fermion combined quantum libraries facilitate the implementation, testing and overall development of variational quantum algorithms dedicated to quantum chemistry problems and provide state-of-the-art quantum computing methods (see Table \ref{Open-VQEpackages} which summarizes the GitHub links of the open-source quantum chemistry packages available to the community).
\begin{table}[h!]
\centering
\begin{tabular}{lcccccc}
\rowcolor[HTML]{C0C0C0}
\textit{package$\,$\textbackslash  $\,$ansatz} & UCCSD & k-UpCCGSD & ADAPT & PNO-UpCCGSD & SPA & QPE\\
\hline
adapt-vqe & $\times$ & $\times$ &\cite{ADAPT-Git} & $\times$ & $\times$ & $\times$\\
\rowcolor[HTML]{EFEFEF}
TEQUILA & \cite{kottmann2021tequila} & \cite{pno} & \cite{kottmann2021tequila} & \cite{kottmann2021tequila,kottmann2021reducing} & \cite{kottmann2021tequila,kottmann2022optimized} & $\times$\\
PennyLane & \cite{bergholm2018pennylane,arrazola2021universal} & \cite{PennyLaneAI} & \cite{bergholm2018pennylane,delgado2021variational} &$\times$ &$\times$&$\times$\\
\rowcolor[HTML]{EFEFEF}
OpenFermion & \cite{mcclean2020openfermion} &$\times$ & $\times$& $\times$&$\times$ &$\times$ \\
QEBAB & \cite{QEBAB} & \cite{QEBAB} & \cite{chan2021molecular,QEBAB} & $\times$ & $\times$ & $\times$\\
\rowcolor[HTML]{EFEFEF}
QFORTE & \cite{stair2021qforte} & $\times$ & \cite{stair2021qforte,stair2021simulating} & $\times$ & $\times$ & \cite{stair2021qforte,stair2020multireference} \\
Qiskit & \cite{Qiskit2021} & $\times$ & \cite{Qiskit2021} & $\times$& $\times$ & \cite{QiskitQPE}\\
\rowcolor[HTML]{EFEFEF}
XACC & \cite{mccaskey2020xacc} & $\times$ &$\times$ & $\times$& $\times$&$\times$\\
QDK &\cite{qdk} &$\times$ &$\times$ & $\times$& $\times$ &\cite{qdk}\\
\rowcolor[HTML]{EFEFEF}
InQuanto &\cite{InQuanto} &$\times$ &\cite{InQuanto} & $\times$& $\times$ &$\times$\\
\hline
\hline
\rowcolor[HTML]{EFEFEF}
OpenVQE/myQLM-fermion
 &\cite{openvqe,myqlm-fermion} &\cite{openvqe} &\cite{openvqe} &$\times$ & $\times$& \cite{myqlm-fermion}\\
\end{tabular}
\caption{Available implementations of individual approaches in a list of most recent open sources quantum chemistry simulation packages (Table referred from \cite{anand2021quantum})
including OpenVQE and myQLM-fermion packages. 
Note: PNO-UpCCGSD stands for pair-natural orbitals-UpCCGSD and SPA for Separable Pair Approximation. The $\times$ notation means that the method is NOT implemented in a given package.}
\label{Open-VQEpackages}
\end{table}
We have shown that OpenVQE enables users to construct modules in a few lines of Python code using simple class structures related to the adaptive UCC family ansätze that we reviewed in the article. We also presented how to easily implement efficient circuits related to fermionic and qubit evolutions with the goal of reducing the number of CNOT gates.
Using these modules, we generated benchmarks on a range of molecular computations from 4 to 22 qubits and obtained reliable results that appeared consistent with those previously obtained in reference literature (see \cite{romero2018strategies,Harsha, kuhn,Yordan} for the UCC  and \cite{grimsley2019adaptive,Yordan,tang2021qubit,shkolnikov2021avoiding} for the ADAPT-VQE approaches). It is worth noting that technically the QLM machine is designed to simulate up to 41 Qubits on a large memory classical computer systems.
OpenVQE is thus designed to simplify the researchers' work, providing simple code modules facilitating the construction of new algorithms and ansätze. 
 
Furthermore, OpenVQE benefits from the myQLM-fermion tools and allow to exchange/interact with other software, like quantum chemistry-oriented quantum computing packages (see Figure.S1 in the Supplementary Material\cite{supplementary}), thus enhancing the community's capacity to work collectively and faster.
We also intend to forge an efficient module that could merge the few keys tools required to run jobs on quantum computers (IBM, Rigetti etc...). For example, Figure.S2 show examples of simple code structure classes that involve the interoperability between different open source packages and QPUs.



We plan to further extend  OpenVQE/myQLM-fermion tools towards reducing the quantum resources by building new modules describing functions that enable:
(i) to taper off the number of qubits in VQE (alongside  active space selection) through using penalty functions\cite{kuroiwa2021penalty}, using point-group symmetry in small molecules \cite{setia2020reducing} and in large molecules\cite{cao2021towards} and studying new functions that could preserve spatial symmetry of molecules (rotational and translational);
(ii) to reduce the number of CNOT gates  (alongside  with MP2 guesses, efficient circuits, fermionic and qubit-ADAPT-VQE) through implementing other adaptive algorithms such as permute qubit-VQE\cite{tkachenko2021correlation}.

Concerning chemical accuracy, we want to implement further methods to go beyond the chemically inspired UCC ansätze described in \cite{anand2021quantum} (i.e. the implemented UCCSD, UCCGSD, spin-complement-gsd, k-UpCCGSD etc...). 
Equally we are interested in other types of UCC ansätze that involve higher-order correlation effects,
such as MP3 or MP4, or triple excitation UCCSD(T), which could be important for strongly correlated systems and could further improve the accuracy as mentioned earlier \cite{lee2018generalized}.
We are also motivated to implement additional algorithms going beyond VQE, such as VQD (ADAPT-VQD) and QITE (ADAPT-QITE) that allow to determine excited state energies \cite{ollitrault2020quantum,Greene,Chan,ville2021leveraging,kamakari2022digital}. 

Concerning optimization effects in the UCC family of methods, since we have demonstrated in the present work that some convergence problems might appear in ADAPT-VQE when targeting certain levels of accuracy, we want to make use of more efficient optimizer methods to overcome this problem.
Of course, the practical use of VQE algorithms on present quantum computers will require to make Darwinian choices in link with qubit/operator counts and circuits depths restrictions: not all algorithms will "make the cut" to the real quantum computing world.
We believe that the present OpenVQE/myQLM-Fermion approach will help the community to study and design new accurate VQE algorithmic "champions" able to efficiently run on present and near-future NISQ quantum computers towards performing accurate quantum chemical computations on complex molecular systems.  
\section*{Acknowledgements}
MR is grateful to the European Union for the  Horizon
2020 (H2020) research grant within the $\langle$NE|AS|QC$\rangle$ project (grant agreement No 951821).  This work has also received funding from the European Research Council (ERC) under the European Union's Horizon 2020 research and innovation program (grant agreement No 810367), project EMC2 (JPP, YM).\\
Support from the PEPR EPiQ and HQI programs is acknowledged.
We would like to thank TotalEnergies for providing HPC ressources. \\
Special thanks also to Diata Traore (LCT) for discussions on the classical quantum chemistry computations; to Baptiste Anselme Martin (TotalEnergies) and Cesar Feniou (LCT/Qubit Pharmaceuticals) for helpful quantum computing discussions. 

\section*{Further Reading}
For further information, we encourage the reader to visit the documentation of the packages: myQLM\cite{myQLM} and myQLM-fermion\cite{myqlm-fermion} and OpenVQE\cite{openvqe}. They cover many subjects discussed here and the tutorials are detailed step-by-step for the user. 
\bibliography{bibliography}

\begin{thebibliography}{100}
\providecommand \doibase [0]{http://dx.doi.org/}%

\bibitem{aspuru2005simulated}
Aspuru-Guzik A, Dutoi AD, Love PJ, Head-Gordon M. Simulated quantum computation
  of molecular energies. {\it Science} 2005\string; 309(5741)\string:
  1704--1707.

\bibitem{bassman2021simulating}
Bassman L, Urbanek M, Metcalf M, Carter J, Kemper AF, Jong dWA. Simulating
  quantum materials with digital quantum computers. {\it Quantum Science and
  Technology} 2021\string; 6(4)\string: 043002.

\bibitem{lehtola2017cluster}
Lehtola S, Tubman NM, Whaley KB, Head-Gordon M. Cluster decomposition of full
  configuration interaction wave functions: A tool for chemical interpretation
  of systems with strong correlation. {\it The Journal of chemical physics}
  2017\string; 147(15)\string: 154105.

\bibitem{gan2006lowest}
Gan Z, Grant DJ, Harrison RJ, Dixon DA. The lowest energy states of the
  group-IIIA--group-VA heteronuclear diatomics: BN, BP, AlN, and AlP from full
  configuration interaction calculations. {\it The Journal of chemical physics}
  2006\string; 125(12)\string: 124311.

\bibitem{vogiatzis2017pushing}
Vogiatzis KD, Ma D, Olsen J, Gagliardi L, De~Jong WA. Pushing
  configuration-interaction to the limit: Towards massively parallel MCSCF
  calculations. {\it The Journal of chemical physics} 2017\string;
  147(18)\string: 184111.

\bibitem{hohenberg1964inhomogeneous}
Hohenberg P, Kohn W. Inhomogeneous electron gas. {\it Physical review}
  1964\string; 136(3B)\string: B864.

\bibitem{kohn1965self}
Kohn W, Sham LJ. Self-consistent equations including exchange and correlation
  effects. {\it Physical review} 1965\string; 140(4A)\string: A1133.

\bibitem{schutt2018machine}
Sch\"utt O, VandeVondele J. Machine learning adaptive basis sets for efficient
  large scale density functional theory simulation. {\it Journal of chemical
  theory and computation} 2018\string; 14(8)\string: 4168--4175.

\bibitem{bartlett1989alternative}
Bartlett RJ, Kucharski SA, Noga J. Alternative coupled-cluster ans{\"a}tze II.
  The unitary coupled-cluster method. {\it Chemical physics letters}
  1989\string; 155(1)\string: 133--140.

\bibitem{kutzelnigg1991error}
Kutzelnigg W. Error analysis and improvements of coupled-cluster theory. {\it
  Theoretica chimica acta} 1991\string; 80(4)\string: 349--386.

\bibitem{harrison1991approximating}
Harrison RJ. Approximating full configuration interaction with selected
  configuration interaction and perturbation theory. {\it The Journal of
  chemical physics} 1991\string; 94(7)\string: 5021--5031.

\bibitem{olsen1996full}
Olsen J, Jo/rgensen P, Koch H, Balkova A, Bartlett RJ. Full
  configuration--interaction and state of the art correlation calculations on
  water in a valence double-zeta basis with polarization functions. {\it The
  Journal of chemical physics} 1996\string; 104(20)\string: 8007--8015.

\bibitem{peris1999perturbatively}
Peris G, Planelles J, Malrieu JP, Paldus J. Perturbatively selected CI as an
  optimal source for externally corrected CCSD. {\it The Journal of chemical
  physics} 1999\string; 110(24)\string: 11708--11716.

\bibitem{taube2006new}
Taube AG, Bartlett RJ. New perspectives on unitary coupled-cluster theory. {\it
  International journal of quantum chemistry} 2006\string; 106(15)\string:
  3393--3401.

\bibitem{bartlett2007coupled}
Bartlett RJ, Musia{\l} M. Coupled-cluster theory in quantum chemistry. {\it
  Reviews of Modern Physics} 2007\string; 79(1)\string: 291.

\bibitem{schriber2016communication}
Schriber JB, Evangelista FA. Communication: An adaptive configuration
  interaction approach for strongly correlated electrons with tunable accuracy.
  {\it The Journal of chemical physics} 2016\string; 144(16)\string: 161106.

\bibitem{nagy2019approaching}
Nagy PR, K{\'a}llay M. Approaching the basis set limit of CCSD (T) energies for
  large molecules with local natural orbital coupled-cluster methods. {\it
  Journal of Chemical Theory and Computation} 2019\string; 15(10)\string:
  5275--5298.

\bibitem{feynman2018simulating}
Feynman RP. Simulating physics with computers. In: CRC Press.  2018 (pp.
  133--153).

\bibitem{benioff1980computer}
Benioff P. The computer as a physical system: A microscopic quantum mechanical
  Hamiltonian model of computers as represented by Turing machines. {\it
  Journal of statistical physics} 1980\string; 22(5)\string: 563--591.

\bibitem{reiher2017elucidating}
Reiher M, Wiebe N, Svore KM, Wecker D, Troyer M. Elucidating reaction
  mechanisms on quantum computers. {\it Proceedings of the national academy of
  sciences} 2017\string; 114(29)\string: 7555--7560.

\bibitem{mcardle2020quantum}
McArdle S, Endo S, Aspuru-Guzik A, Benjamin SC, Yuan X. Quantum computational
  chemistry. {\it Reviews of Modern Physics} 2020\string; 92(1)\string: 015003.

\bibitem{cao2019quantum}
Cao Y, Romero J, Olson JP, et al. Quantum chemistry in the age of quantum
  computing. {\it Chemical reviews} 2019\string; 119(19)\string: 10856--10915.

\bibitem{bauer2020quantum}
Bauer B, Bravyi S, Motta M, Chan GKL. Quantum algorithms for quantum chemistry
  and quantum materials science. {\it Chemical Reviews} 2020\string;
  120(22)\string: 12685--12717.

\bibitem{yeter2021benchmarking}
Yeter-Aydeniz K, Gard BT, Jakowski J, et al. Benchmarking quantum chemistry
  computations with variational, imaginary time evolution, and Krylov space
  solver algorithms. {\it Advanced Quantum Technologies} 2021\string;
  4(7)\string: 2100012.

\bibitem{whitfield2011simulation}
Whitfield JD, Biamonte J, Aspuru-Guzik A. Simulation of electronic structure
  Hamiltonians using quantum computers. {\it Molecular Physics} 2011\string;
  109(5)\string: 735--750.

\bibitem{preskill2018quantum}
Preskill J. Quantum computing in the NISQ era and beyond. {\it Quantum}
  2018\string; 2\string: 79.

\bibitem{elfving2020will}
Elfving VE, Broer BW, Webber M, et al. How will quantum computers provide an
  industrially relevant computational advantage in quantum chemistry?. {\it
  arXiv preprint arXiv:2009.12472} 2020.

\bibitem{bharti2022noisy}
Bharti K, Cervera-Lierta A, Kyaw TH, et al. Noisy intermediate-scale quantum
  algorithms. {\it Reviews of Modern Physics} 2022\string; 94(1)\string:
  015004.

\bibitem{peruzzo2014variational}
Peruzzo A, McClean J, Shadbolt P, et al. A variational eigenvalue solver on a
  photonic quantum processor. {\it Nature communications} 2014\string;
  5(1)\string: 1--7.

\bibitem{mcclean2016theory}
McClean JR, Romero J, Babbush R, Aspuru-Guzik A. The theory of variational
  hybrid quantum-classical algorithms. {\it New Journal of Physics}
  2016\string; 18(2)\string: 023023.

\bibitem{cerezo2021variational}
Cerezo M, Arrasmith A, Babbush R, et al. Variational quantum algorithms. {\it
  Nature Reviews Physics} 2021\string; 3(9)\string: 625--644.

\bibitem{google2020hartree}
Quantum GA, Collaborators*† , Arute F, et al. Hartree-Fock on a
  superconducting qubit quantum computer. {\it Science} 2020\string;
  369(6507)\string: 1084--1089.

\bibitem{kandala2017hardware}
Kandala A, Mezzacapo A, Temme K, et al. Hardware-efficient variational quantum
  eigensolver for small molecules and quantum magnets. {\it Nature}
  2017\string; 549(7671)\string: 242--246.

\bibitem{hempel2018quantum}
Hempel C, Maier C, Romero J, et al. Quantum chemistry calculations on a
  trapped-ion quantum simulator. {\it Physical Review X} 2018\string;
  8(3)\string: 031022.

\bibitem{romero2018strategies}
Romero J, Babbush R, McClean JR, Hempel C, Love PJ, Aspuru-Guzik A. Strategies
  for quantum computing molecular energies using the unitary coupled cluster
  ansatz. {\it Quantum Science and Technology} 2018\string; 4(1)\string:
  014008.

\bibitem{anand2021quantum}
Anand A, Schleich P, Alperin-Lea S, et al. A quantum computing view on unitary
  coupled cluster theory. {\it Chemical Society Reviews} 2022.

\bibitem{gard2020efficient}
Gard BT, Zhu L, Barron GS, Mayhall NJ, Economou SE, Barnes E. Efficient
  symmetry-preserving state preparation circuits for the variational quantum
  eigensolver algorithm. {\it npj Quantum Information} 2020\string;
  6(1)\string: 1--9.

\bibitem{ryabinkin2018qubit}
Ryabinkin IG, Yen TC, Genin SN, Izmaylov AF. Qubit coupled cluster method: a
  systematic approach to quantum chemistry on a quantum computer. {\it Journal
  of chemical theory and computation} 2018\string; 14(12)\string: 6317--6326.

\bibitem{kim2017robust}
Kim IH, Swingle B. Robust entanglement renormalization on a noisy quantum
  computer. {\it arXiv preprint arXiv:1711.07500} 2017.

\bibitem{wecker2015progress}
Wecker D, Hastings MB, Troyer M. Progress towards practical quantum variational
  algorithms. {\it Physical Review A} 2015\string; 92(4)\string: 042303.

\bibitem{grimsley2019adaptive}
Grimsley HR, Economou SE, Barnes E, Mayhall NJ. An adaptive variational
  algorithm for exact molecular simulations on a quantum computer. {\it Nature
  communications} 2019\string; 10(1)\string: 1--9.

\bibitem{mcclean2017hybrid}
McClean JR, Kimchi-Schwartz ME, Carter J, De~Jong WA. Hybrid quantum-classical
  hierarchy for mitigation of decoherence and determination of excited states.
  {\it Physical Review A} 2017\string; 95(4)\string: 042308.

\bibitem{myQLM}
myQLM package. \url{https://myqlm.github.io/}{https://myqlm.github.io/}; .

\bibitem{QLMbinders}
Interoperability with myQLM.
  \url{https://myqlm.github.io/myqlm_specific/interoperability.html}{https://myqlm.github.io/\textit{myqlm\_specific}/interoperability.html};
  .

\bibitem{wille2019ibm}
Wille R, Van~Meter R, Naveh Y. IBM’s Qiskit tool chain: Working with and
  developing for real quantum computers. In: IEEE. ; 2019\string: 1234--1240.

\bibitem{cirqql}
Developers C. Cirq.
  \url{https://github.com/quantumlib/Cirq/graphs/contributors};  2021.
\newblock {See full list of authors on Github:
  https://github.com/quantumlib/Cirq/graphs/contributors}

\bibitem{smith2016practical}
Smith RS, Curtis MJ, Zeng WJ. A practical quantum instruction set architecture.
  {\it arXiv preprint arXiv:1608.03355} 2016.

\bibitem{nooijen2000can}
Nooijen M. Can the eigenstates of a many-body hamiltonian be represented
  exactly using a general two-body cluster expansion?. {\it Physical review
  letters} 2000\string; 84(10)\string: 2108.

\bibitem{lee2018generalized}
Lee J, Huggins WJ, Head-Gordon M, Whaley KB. Generalized unitary coupled
  cluster wave functions for quantum computation. {\it Journal of chemical
  theory and computation} 2018\string; 15(1)\string: 311--324.

\bibitem{tang2021qubit}
Tang HL, Shkolnikov V, Barron GS, et al. qubit-adapt-vqe: An adaptive algorithm
  for constructing hardware-efficient ans{\"a}tze on a quantum processor. {\it
  PRX Quantum} 2021\string; 2(2)\string: 020310.

\bibitem{Xia}
Xia R, Kais S. Qubit coupled cluster singles and doubles variational quantum
  eigensolver ansatz for electronic structure calculations. {\it Quantum
  Science and Technology} 2020\string; 6(1)\string: 015001.

\bibitem{shkolnikov2021avoiding}
Shkolnikov V, Mayhall NJ, Economou SE, Barnes E. Avoiding symmetry roadblocks
  and minimizing the measurement overhead of adaptive variational quantum
  eigensolvers. {\it arXiv preprint arXiv:2109.05340} 2021.

\bibitem{openvqe}
Open-VQE package.
  \url{https://github.com/OpenVQE/OpenVQE.git}\url{https://openvqe.github.io/OpenVQE/}{Repository:
  https://github.com/OpenVQE/OpenVQE.git and
  documentation:https://openvqe.github.io/OpenVQE/}; .

\bibitem{myqlm-fermion}
myqlm-fermion.
  \url{https://github.com/myQLM/myqlm-fermion}\url{https://myqlm.github.io/qat-fermion.html}{Repository:
  https://github.com/myQLM/myqlm-fermion and documentation:
  https://myqlm.github.io/qat-fermion.html}; .

\bibitem{Cizek1966}
{\v{C}}{\'\i}{\v{z}}ek J. On the correlation problem in atomic and molecular
  systems. Calculation of wavefunction components in Ursell-type expansion
  using quantum-field theoretical methods. {\it The Journal of Chemical
  Physics} 1966\string; 45(11)\string: 4256--4266.

\bibitem{Cizek1980}
Cizek J, Paldus J. Coupled cluster approach. {\it Physica Scripta} 1980\string;
  21(3-4)\string: 251.

\bibitem{Yung2014}
Yung MH, Casanova J, Mezzacapo A, et al. From transistor to trapped-ion
  computers for quantum chemistry. {\it Scientific reports} 2014\string;
  4(1)\string: 1--7.

\bibitem{barkoutsos2018quantum}
Barkoutsos PK, Gonthier JF, Sokolov I, et al. Quantum algorithms for electronic
  structure calculations: Particle-hole Hamiltonian and optimized wave-function
  expansions. {\it Physical Review A} 2018\string; 98(2)\string: 022322.

\bibitem{ganzhorn2019gate}
Ganzhorn M, Egger DJ, Barkoutsos P, et al. Gate-efficient simulation of
  molecular eigenstates on a quantum computer. {\it Physical Review Applied}
  2019\string; 11(4)\string: 044092.

\bibitem{sennane2022calculating}
Sennane W, Piquemal JP, Ran{\v{c}}i{\'c} MJ. Calculating the ground state
  energy of benzene under spatial deformations with noisy quantum computing.
  {\it arXiv preprint arXiv:2203.05275} 2022.

\bibitem{mcclean2018barren}
McClean JR, Boixo S, Smelyanskiy VN, Babbush R, Neven H. Barren plateaus in
  quantum neural network training landscapes. {\it Nature communications}
  2018\string; 9(1)\string: 1--6.

\bibitem{cerezo2021cost}
Cerezo M, Sone A, Volkoff T, Cincio L, Coles PJ. Cost function dependent barren
  plateaus in shallow parametrized quantum circuits. {\it Nature
  communications} 2021\string; 12(1)\string: 1--12.

\bibitem{Taube2006}
Taube AG, Bartlett RJ. New perspectives on unitary coupled-cluster theory. {\it
  International journal of quantum chemistry} 2006\string; 106(15)\string:
  3393--3401.

\bibitem{shen2017quantum}
Shen Y, Zhang X, Zhang S, Zhang JN, Yung MH, Kim K. Quantum implementation of
  the unitary coupled cluster for simulating molecular electronic structure.
  {\it Physical Review A} 2017\string; 95(2)\string: 020501.

\bibitem{fradkin1989jordan}
Fradkin E. Jordan-Wigner transformation for quantum-spin systems in two
  dimensions and fractional statistics. {\it Physical review letters}
  1989\string; 63(3)\string: 322.

\bibitem{Evangelista2019}
Evangelista FA, Chan GKL, Scuseria GE. Exact parameterization of fermionic wave
  functions via unitary coupled cluster theory. {\it The Journal of chemical
  physics} 2019\string; 151(24)\string: 244112.

\bibitem{Grimsleytrotter}
Grimsley HR, Claudino D, Economou SE, Barnes E, Mayhall NJ. Is the trotterized
  uccsd ansatz chemically well-defined?. {\it Journal of chemical theory and
  computation} 2019\string; 16(1)\string: 1--6.

\bibitem{hatano2005finding}
Hatano N, Suzuki M. Finding exponential product formulas of higher orders. In:
  Springer.  2005 (pp. 37--68).

\bibitem{Yordan2019}
Yordanov YS, Barnes CH. Implementation of a general single-qubit positive
  operator-valued measure on a circuit-based quantum computer. {\it Physical
  Review A} 2019\string; 100(6)\string: 062317.

\bibitem{Yordan2020}
Yordanov YS, Arvidsson-Shukur DR, Barnes CH. Efficient quantum circuits for
  quantum computational chemistry. {\it Physical Review A} 2020\string;
  102(6)\string: 062612.

\bibitem{Yordan2021}
Yordanov YS, Armaos V, Barnes CH, Arvidsson-Shukur DR. Qubit-excitation-based
  adaptive variational quantum eigensolver. {\it Communications Physics}
  2021\string; 4(1)\string: 1--11.

\bibitem{Yordan}
Yordanov Y. {\it Quantum computational chemistry methods for early-stage
  quantum computers}. PhD thesis. University of Cambridge, Cambridge, UK;
  2021.

\bibitem{William2020}
Huggins WJ, Lee J, Baek U, O’Gorman B, Whaley KB. A non-orthogonal
  variational quantum eigensolver. {\it New Journal of Physics} 2020\string;
  22(7)\string: 073009.

\bibitem{cao2022progress}
Cao C, Hu J, Zhang W, et al. Progress toward larger molecular simulation on a
  quantum computer: Simulating a system with up to 28 qubits accelerated by
  point-group symmetry. {\it Physical Review A} 2022\string; 105(6)\string:
  062452.

\bibitem{Greene}
Greene-Diniz G, Mu{\~n}oz~Ramo D. Generalized unitary coupled cluster
  excitations for multireference molecular states optimized by the variational
  quantum eigensolver. {\it International Journal of Quantum Chemistry}
  2021\string; 121(4)\string: e26352.

\bibitem{Chan}
Chan HHS, Fitzpatrick N, Segarra-Mart{\'\i} J, Bearpark MJ, Tew DP. Molecular
  excited state calculations with adaptive wavefunctions on a quantum
  eigensolver emulation: reducing circuit depth and separating spin states.
  {\it Physical Chemistry Chemical Physics} 2021\string; 23(46)\string:
  26438--26450.

\bibitem{Arthur2019}
Rattew AG, Hu S, Pistoia M, Chen CFR, Wood S. A domain-agnostic,
  noise-resistant evolutionary variational quantum eigensolver for
  hardware-efficient optimization in the Hilbert space. {\it DeepAI} 2019.

\bibitem{Ilya2020}
Ryabinkin IG, Lang RA, Genin SN, Izmaylov AF. Iterative qubit coupled cluster
  approach with efficient screening of generators. {\it Journal of chemical
  theory and computation} 2020\string; 16(2)\string: 1055--1063.

\bibitem{Robert2020}
Lang RA, Ryabinkin IG, Izmaylov AF. Unitary transformation of the electronic
  hamiltonian with an exact quadratic truncation of the
  baker-campbell-hausdorff expansion. {\it Journal of Chemical Theory and
  Computation} 2020\string; 17(1)\string: 66--78.

\bibitem{SukinSim}
Sim S, Romero J, Gonthier JF, Kunitsa AA. Adaptive pruning-based optimization
  of parameterized quantum circuits. {\it Quantum Science and Technology}
  2021\string; 6(2)\string: 025019.

\bibitem{JieLiu}
Liu J, Li Z, Yang J. An efficient adaptive variational quantum solver of the
  Schr{\"o}dinger equation based on reduced density matrices. {\it The Journal
  of chemical physics} 2021\string; 154(24)\string: 244112.

\bibitem{supplementary}
Haidar M, Rančić MJ, Ayral T, Maday Y, Piquemal JP. See Supplemental Material
  for openVQE package: interoperability discussion and numerical results..

\bibitem{atosqlm}
Quantum Learning Machine.
  \url{https://atos.net/en/solutions/quantum-learning-machine}{ATOS Quantum
  Learning Machine}; .

\bibitem{Martiel2020a}
Martiel S, Brugière dTG. Architecture aware compilation of quantum circuits
  via lazy synthesis. {\it Quantum} 2020.
\newblock \href {\doibase 10.48550/ARXIV.2012.09663} {doi:
  10.48550/ARXIV.2012.09663}

\bibitem{Brugiere2021}
De~Brugi\`{e}re TG, Baboulin M, Valiron B, Martiel S, Allouche C. Gaussian
  Elimination versus Greedy Methods for the Synthesis of Linear Reversible
  Circuits. {\it ACM Transactions on Quantum Computing} 2021\string; 2(3).
\newblock \href {\doibase 10.1145/3474226} {doi: 10.1145/3474226}

\bibitem{Vandaele2021}
Vandaele V, Martiel S, Brugière G.~dT. Phase polynomials synthesis algorithms
  for NISQ architectures and beyond. {\it Quantum Science and Technology} 2022.

\bibitem{DeBrugiere2022}
Brugière TGd, Baboulin M, Valiron B, Martiel S, Allouche C. Reducing the Depth
  of Linear Reversible Quantum Circuits. {\it IEEE Transactions on Quantum
  Engineering} 2021\string; 2\string: 1-22.
\newblock \href {\doibase 10.1109/TQE.2021.3091648} {doi:
  10.1109/TQE.2021.3091648}

\bibitem{DeBrugiere2022a}
Brugi{\`{e}}re dTG, Baboulin M, Valiron B, Martiel S, Allouche C. {Decoding
  techniques applied to the compilation of CNOT circuits for NISQ
  architectures}. {\it ScienceDirect} 2022\string; 214(February 2022)\string:
  1--31.
\newblock \href {\doibase 10.1016/j.scico.2021.102726} {doi:
  10.1016/j.scico.2021.102726}

\bibitem{Vidal2003}
Vidal G. Efficient Classical Simulation of Slightly Entangled Quantum
  Computations. {\it Phys. Rev. Lett.} 2003\string; 91\string: 147902.
\newblock \href {\doibase 10.1103/PhysRevLett.91.147902} {doi:
  10.1103/PhysRevLett.91.147902}

\bibitem{Rudiak-Gould2006}
Rudiak-Gould B. The sum-over-histories formulation of quantum computing. {\it
  arXiv preprint quant-ph/0607151} 2006.

\bibitem{Miller2006}
Miller DM, Thornton MA. QMDD: A decision diagram structure for reversible and
  quantum circuits. In: IEEE. ; 2006\string: 30--30.

\bibitem{2020SciPy-NMeth}
Virtanen P, Gommers R, Oliphant TE, et al. {{SciPy} 1.0: Fundamental Algorithms
  for Scientific Computing in Python}. {\it Nature Methods} 2020\string;
  17\string: 261--272.
\newblock \href {\doibase 10.1038/s41592-019-0686-2} {doi:
  10.1038/s41592-019-0686-2}

\bibitem{sun2018pyscf}
Sun Q, Berkelbach TC, Blunt NS, et al. PySCF: the Python-based simulations of
  chemistry framework. {\it Wiley Interdisciplinary Reviews: Computational
  Molecular Science} 2018\string; 8(1)\string: e1340.

\bibitem{scipy}
Virtanen P, Gommers R, Oliphant TE, et al. {{SciPy} 1.0: Fundamental Algorithms
  for Scientific Computing in Python}. {\it Nature Methods} 2020\string;
  17\string: 261--272.
\newblock \href {\doibase 10.1038/s41592-019-0686-2} {doi:
  10.1038/s41592-019-0686-2}

\bibitem{gunnarsson1976exchange}
Gunnarsson O, Lundqvist BI. Exchange and correlation in atoms, molecules, and
  solids by the spin-density-functional formalism. {\it Physical Review B}
  1976\string; 13(10)\string: 4274.

\bibitem{ADAPT-Git}
ADAPT package.
  \url{https://github.com/mayhallgroup/adapt-vqe.git}{https://github.com/mayhallgroup/adapt-vqe.git};
  .

\bibitem{kottmann2021tequila}
Kottmann JS, Alperin-Lea S, Tamayo-Mendoza T, et al. Tequila: A platform for
  rapid development of quantum algorithms. {\it Quantum Science and Technology}
  2021\string; 6(2)\string: 024009.

\bibitem{pno}
Pair-natural orbitals in tequila package.
  \url{https://github.com/aspuru-guzik-group/tequila}{https://github.com/aspuru-guzik-group/tequila};
  .

\bibitem{kottmann2021reducing}
Kottmann JS, Schleich P, Tamayo-Mendoza T, Aspuru-Guzik A. Reducing qubit
  requirements while maintaining numerical precision for the variational
  quantum eigensolver: A basis-set-free approach. {\it The Journal of Physical
  Chemistry Letters} 2021\string; 12(1)\string: 663--673.

\bibitem{kottmann2022optimized}
Kottmann JS, Aspuru-Guzik A. Optimized low-depth quantum circuits for molecular
  electronic structure using a separable-pair approximation. {\it Physical
  Review A} 2022\string; 105(3)\string: 032449.

\bibitem{bergholm2018pennylane}
Bergholm V, Izaac J, Schuld M, et al. Pennylane: Automatic differentiation of
  hybrid quantum-classical computations. {\it arXiv preprint arXiv:1811.04968}
  2018.

\bibitem{arrazola2021universal}
Arrazola JM, Di~Matteo O, Quesada N, Jahangiri S, Delgado A, Killoran N.
  Universal quantum circuits for quantum chemistry. {\it arXiv preprint
  arXiv:2106.13839} 2021.

\bibitem{PennyLaneAI}
PennyLaneAI package.
  \url{https://github.com/PennyLaneAI/pennylane.git}{https://github.com/PennyLaneAI/pennylane.git};
  .

\bibitem{delgado2021variational}
Delgado A, Arrazola JM, Jahangiri S, et al. Variational quantum algorithm for
  molecular geometry optimization. {\it Physical Review A} 2021\string;
  104(5)\string: 052402.

\bibitem{mcclean2020openfermion}
McClean JR, Rubin NC, Sung KJ, et al. OpenFermion: the electronic structure
  package for quantum computers. {\it Quantum Science and Technology}
  2020\string; 5(3)\string: 034014.

\bibitem{QEBAB}
QEBAB package.
  \url{https://github.com/hanschanhs/QEBAB.git}{https://github.com/hanschanhs/QEBAB.git};
  .

\bibitem{chan2021molecular}
Chan HHS, Fitzpatrick N, Segarra-Mart{\'\i} J, Bearpark MJ, Tew DP. Molecular
  excited state calculations with adaptive wavefunctions on a quantum
  eigensolver emulation: reducing circuit depth and separating spin states.
  {\it Physical Chemistry Chemical Physics} 2021\string; 23(46)\string:
  26438--26450.

\bibitem{stair2021qforte}
Stair NH, Evangelista FA. Qforte: an efficient state simulator and quantum
  algorithms library for molecular electronic structure. {\it arXiv preprint
  arXiv:2108.04413} 2021.

\bibitem{stair2021simulating}
Stair NH, Evangelista FA. Simulating many-body systems with a projective
  quantum eigensolver. {\it PRX Quantum} 2021\string; 2(3)\string: 030301.

\bibitem{stair2020multireference}
Stair NH, Huang R, Evangelista FA. A multireference quantum krylov algorithm
  for strongly correlated electrons. {\it Journal of chemical theory and
  computation} 2020\string; 16(4)\string: 2236--2245.

\bibitem{Qiskit2021}
Aleksandrowicz G, Alexander T, Barkoutsos P, et al. {Qiskit: An Open-source
  Framework for Quantum Computing}.
  \url{https://doi.org/10.5281/zenodo.2562111};  2019

\bibitem{QiskitQPE}
Quantum phase estimation algorithm in Qiskit package.
  \url{https://qiskit.org/textbook/ch-algorithms/quantum-phase-estimation.html}{https://qiskit.org/textbook/ch-algorithms/quantum-phase-estimation.html};
  .

\bibitem{mccaskey2020xacc}
McCaskey AJ, Lyakh DI, Dumitrescu EF, Powers SS, Humble TS. XACC: a
  system-level software infrastructure for heterogeneous quantum--classical
  computing. {\it Quantum Science and Technology} 2020\string; 5(2)\string:
  024002.

\bibitem{qdk}
Quantum Development Kit.
  \url{https://github.com/microsoft/quantum.git}{https://github.com/microsoft/quantum.git};
  .

\bibitem{InQuanto}
InQuanto package.
  \url{https://medium.com/cambridge-quantum-computing/introduction-to-the-inquanto-computational-chemistry-platform-for-quantum-computers}{https://medium.com/cambridge-quantum-computing/introduction-to-the-inquanto-computational-chemistry-platform-for-quantum-computers};
  .

\bibitem{Harsha}
Harsha G, Shiozaki T, Scuseria GE. On the difference between variational and
  unitary coupled cluster theories. {\it The Journal of chemical physics}
  2018\string; 148(4)\string: 044107.

\bibitem{kuhn}
K\"uhn M, Zanker S, Deglmann P, Marthaler M, Wei{\ss} H. Accuracy and resource
  estimations for quantum chemistry on a near-term quantum computer. {\it
  Journal of chemical theory and computation} 2019\string; 15(9)\string:
  4764--4780.

\bibitem{kuroiwa2021penalty}
Kuroiwa K, Nakagawa YO. Penalty methods for a variational quantum eigensolver.
  {\it Physical Review Research} 2021\string; 3(1)\string: 013197.

\bibitem{setia2020reducing}
Setia K, Chen R, Rice JE, Mezzacapo A, Pistoia M, Whitfield JD. Reducing qubit
  requirements for quantum simulations using molecular point group symmetries.
  {\it Journal of Chemical Theory and Computation} 2020\string; 16(10)\string:
  6091--6097.

\bibitem{cao2021towards}
Cao C, Hu J, Zhang W, et al. Towards a Larger Molecular Simulation on the
  Quantum Computer: Up to 28 Qubits Systems Accelerated by Point Group
  Symmetry. {\it arXiv preprint arXiv:2109.02110} 2021.

\bibitem{tkachenko2021correlation}
Tkachenko NV, Sud J, Zhang Y, et al. Correlation-informed permutation of qubits
  for reducing ansatz depth in the variational quantum eigensolver. {\it PRX
  Quantum} 2021\string; 2(2)\string: 020337.

\bibitem{ollitrault2020quantum}
Ollitrault PJ, Kandala A, Chen CF, et al. Quantum equation of motion for
  computing molecular excitation energies on a noisy quantum processor. {\it
  Physical Review Research} 2020\string; 2(4)\string: 043140.

\bibitem{ville2021leveraging}
Ville JL, Morvan A, Hashim A, et al. Leveraging randomized compiling for the
  QITE algorithm. {\it arXiv preprint arXiv:2104.08785} 2021.

\bibitem{kamakari2022digital}
Kamakari H, Sun SN, Motta M, Minnich AJ. Digital quantum simulation of open
  quantum systems using quantum imaginary--time evolution. {\it PRX Quantum}
  2022\string; 3(1)\string: 010320.

\end{thebibliography}

\end{document}